\documentclass[12pt]{article}
\usepackage{graphicx}
\usepackage{amsfonts}
\usepackage{amsmath,amssymb}
\usepackage{mathrsfs}
\usepackage{setspace}
\usepackage{cite}

\def\beq{\begin{equation}}
\def\eeq{\end{equation}}
\def\bea{\begin{eqnarray}}
\def\eea{\end{eqnarray}}
\def\noi{\noindent}
\def\non{\nonumber}
\def\bib{\bibitem}
\def\ba{\begin{array}}
\def\ea{\end{array}}

\def\de{\delta}

\begin{document}

\begin{center}
{\Large \bf \sf Clusters of bound particles in the derivative 
$\de$-function Bose gas}

\vspace{1.3cm}

{\sf B. Basu-Mallick$^1$\footnote{e-mail address:
bireswar.basumallick@saha.ac.in},
Tanaya Bhattacharyya$^2$\footnote{e-mail address:
tanaya.bhattacharyya@googlemail.com}
and Diptiman Sen$^3$\footnote{e-mail address: diptiman@cts.iisc.ernet.in}}
\bigskip

{\em $^1$Theory Division, Saha Institute of Nuclear Physics, \\
1/AF Bidhan Nagar, Kolkata 700 064, India}

\bigskip

{\em $^2$Department of Physics, St. Xavier's College, \\
30 Park Street, Kolkata 700 016, India}

\bigskip

{\em $^3$Centre for High Energy Physics, Indian Institute of Science, \\
Bangalore 560 012, India}
\end{center}

\bigskip
\bigskip

\noi {\bf Abstract}

In this paper we discuss a novel procedure for constructing clusters 
of bound particles in the case of a quantum integrable derivative 
$\de$-function Bose gas in one dimension. It is shown that clusters of bound 
particles can be constructed for this Bose gas for some special values of the 
coupling constant, by taking the quasi-momenta associated with the 
corresponding Bethe state to be equidistant points on a {\it single} circle in
the complex momentum plane. We also establish a connection between these 
special values of the coupling constant and some fractions belonging to the 
Farey sequences in number theory. This connection leads to a classification of
the clusters of bound particles associated with the derivative $\de$-function 
Bose gas and allows us to study various properties of these clusters like 
their size and their stability under the variation of the coupling constant. 

\newpage

\noi \section{Introduction}
\renewcommand{\theequation}{1.{\arabic{equation}}}
\setcounter{equation}{0}
\medskip
 
Exact solutions of one-dimensional (1D) quantum integrable many-body systems 
with short range interactions have emerged as an active area of research 
\cite{LL63,Mc64,Ya68,YY69,Th81,Fa80,Sk90,KBI93,Ta99,Ol98,GS03,CC07,
CC07a,Do10,PS11,KMT09,KMT10,MF10,KCI11,CCGOR11 }, 
due to their effectiveness in describing recent experiments using strongly 
interacting ultracold atomic gases 
\cite{K02,SPTH02,KGDS03,KWW04,Pa04,KWW05,KWW06,Am08}.
The nearly 1D motion of ultracold bosons 
is achieved in such experiments by confining the bosons within waveguides 
that tightly trap their motion in the two transverse directions,
and allow them to move only in the third direction. Many results of these 
experiments can be understood within the framework of the well known 
Lieb-Liniger model or the $\de$-function Bose gas; this is 
a 1D quantum integrable system with the Hamiltonian for $N$ particles given by 
\beq H_N = -\hbar^2\sum_{j=1}^N\frac{\partial^2}{\partial x_j^2} 
+2\hbar^2 \mu ~\sum_{l<m} \de (x_l - x_m) \, , \label{a1} \eeq
where $\mu$ is the coupling constant. Exact eigenfunctions of the Hamiltonian 
(\ref{a1}) are constructed by using the methods of coordinate as well as 
algebraic Bethe ansatz \cite{LL63,Mc64,Ya68,YY69,Th81,Fa80,Sk90,KBI93,Ta99}.
Moreover, the equilibrium properties of this system have been
studied by employing the thermodynamic Bethe ansatz and various correlation 
functions have been computed through different approaches
\cite{YY69,KBI93,Ta99,Ol98,GS03,CC07,CC07a,Do10,PS11,KMT09,KMT10,MF10,
KCI11,CCGOR11}. 

In this context it may be recalled that, for a class of exactly solvable 
dynamical systems, the method of coordinate Bethe ansatz directly yields 
the eigenfunctions in the coordinate representation. One can study the 
asymptotic form of these eigenfunctions in the limit of infinite length of 
the system (i.e., when all $x_i$'s are allowed to take value
in the range $-\infty < x_i < \infty $). 
If the probability density associated with an eigenfunction decays sufficiently
fast when any of the particle coordinates tends towards infinity (keeping the 
centre of mass coordinate fixed, for a translationally invariant system), 
a bound state is formed. The stability of a bound state, in the 
presence of small external perturbations, can be determined by calculating 
its binding energy. It is well known,
for the case of the $\de$-function Bose gas (\ref{a1}) with $N \geq 2$,
that bound states with positive binding energies exist for all negative values
of the coupling constant $\mu$ \cite{Mc64,Ya68,Th81,Fa80,Sk90,KBI93,Ta99}.
The quasi-momenta associated with such a bound state are represented by 
equidistant points lying on a straight line or `string' parallel to 
the imaginary axis in the complex momentum plane. Moreover,
for the case of the $\de$-function Bose gas with negative values of
the coupling constant, one can construct Bethe eigenfunctions corresponding to
more complex structures like clusters of bound particles and show that those 
clusters of bound particles are stable under scattering. The quasi-momenta 
corresponding to such clusters of bound particles are represented 
through discrete points lying on several `strings', 
all of which are parallel to the imaginary axis in the complex momentum 
plane \cite{Ya68,Ta99,CC07,CC07a,Do10,PS11}.

Similar to the case of the $\de$-function Bose gas mentioned above, 
there exists another exactly solvable and quantum integrable bosonic system 
with a Hamiltonian given by 
\beq {\cal H}_N 
~=~ -\hbar^2 ~\sum_{j=1}^N ~\frac{\partial^2}{\partial x_j^2} ~+~ 2i
\hbar^2 \eta ~\sum_{l<m} ~\de(x_l - x_m )~ \Big( \frac{\partial}{\partial
x_l} + \frac{\partial} {\partial x_m} \Big) \, , \label{a2} \eeq
where $\eta$ is a real nonzero coupling constant 
\cite{Gu87,KB93,SMB94,BB02,BB03}. The Hamiltonian (\ref{a2}) of this 
derivative $\de$-function Bose gas can be obtained by projecting that of
an integrable derivative nonlinear Schr\"{o}dinger (DNLS) quantum
field model on the $N$-particle subspace. 
Classical and quantum versions of such DNLS field
models have found applications in different areas of physics like circularly
polarized nonlinear Alfven waves in plasma, quantum properties of optical
solitons in fibers, and in some chiral Tomonaga-Luttinger liquids obtained 
from the Chern-Simons model defined in two dimensions 
\cite{CLL79,KN78,MP96,WSKI78,Cl92,RS91,AGJPS96,Ra96}.
The scattering and bound states of the derivative $\de$-function 
Bose gas (\ref{a2}) have been studied extensively by
using the methods of coordinate as well as algebraic Bethe ansatz 
\cite{SMB94,Gu87, KB93,BB02,BB03,BBS03,BBS04,BBS04a}. 
It turns out that the quasi-momenta associated with a bound state of this 
model can be represented by equidistant points on a circle or circular 
`string' with arbitrary radius and having its centre at the origin of the 
complex momentum plane. For the cases $N=2$ and $N=3$, this type of bound 
states can be constructed for any value of $\eta$ within its full range: 
$0 < \mid \eta \mid < \infty$. However, for any given value of $N \geq 4$, 
the derivative $\de$-function Bose gas allows bound 
states in only certain non-overlapping ranges of the coupling constant $\eta$
(the union of these ranges yields a proper subset of the full range of $\eta$),
and such non-overlapping ranges of $\eta$ 
can be determined by using the Farey sequences in number theory 
\cite{BBS03,BBS04,BBS04a}. Furthermore, for any given value of $N \geq 3$, 
bound states with positive as well as negative binding energies can be 
constructed. From the above discussions it is evident that the bound states 
of this derivative $\de$-function Bose gas exhibit a much richer structure in 
comparison to the case of the $\de$-function Bose gas. 

The aim of the present work is to explore how 
clusters of bound particles can be constructed in the simplest possible
way for the derivative $\de$-function Bose gas (\ref{a2}).
In analogy with the case of the $\de$-function Bose gas, one may 
think that clusters of bound particles can only be constructed
for the above mentioned case by properly assigning 
the corresponding quasi-momenta on several concentric
circles or circular `strings' in the complex momentum plane.
However in this article we shall show that for the derivative $\de$-function 
Bose gas with some special values of the coupling constant $\eta$, clusters 
of bound particles can be constructed in a much simpler way by assigning 
the corresponding quasi-momenta as equidistant points on a {\it single} 
circle having its centre at the origin of the complex momentum plane.
The arrangement of this article is as follows. 
In Sec. 2, we first briefly review the construction of Bethe eigenstates
for the case of the derivative $\de$-function Bose gas. Then 
we consider a sufficient condition for which a Bethe eigenstate
would represent clusters of bound particles. This sufficient condition for 
obtaining clusters of bound particles has not attracted much attention 
in the literature, probably because it does
not yield any solution at all for the case of the $\de$-function Bose gas. 
However, in Sec. 3, we show that this sufficient condition yields many 
nontrivial solutions for the case of the derivative $\de$-function Bose gas.
Subsequently, by using some properties of the Farey sequence,
we classify all possible solutions of this sufficient 
condition and obtain different types of clusters of bound particles.
In Sec. 4 we discuss various properties of such clusters of bound particles, 
such as the sizes of the clusters, their stability under the variation of the 
coupling constant, and their binding energy. We end with some concluding 
remarks in Sec. 5. 

\noi \section{Conditions for forming clusters of bound particles}
\renewcommand{\theequation}{2.{\arabic{equation}}}
\setcounter{equation}{0}

\medskip

In the coordinate representation, the eigenvalue equation for the Hamiltonian 
(\ref{a2}) of the derivative $\de$-function Bose gas may be written as
\bea {\cal H}_N ~\tau_N( x_1, x_2, \cdots , x_N ) ~=~ E ~\tau_N( x_1, x_2, 
\cdots , x_N )~, \label{b1} \eea
where $\tau_N( x_1, x_2, \cdots , x_N )$ is a completely symmetric
$N$-particle wave function. Since ${\cal H}_N$ commutes with
the total momentum operator given by
\beq {\cal P}_N ~=~ -i\hbar ~\sum_{j=1}^N ~\frac{\partial}{\partial x_j} ~ \, ,
\label{b2} \eeq
$\tau_N( x_1, x_2, \cdots , x_N )$ can be chosen as a simultaneous 
eigenfunction of these two commuting operators. It may be noted that
the Hamiltonian (\ref{a2}) and momentum (\ref{b2}) operators
enjoy the scaling property ${\cal H}_N \rightarrow \lambda^2 {\cal H}_N$ and
${\cal P}_N \rightarrow \lambda {\cal P}_N$, when all the 
coordinates are transformed as $x_i \rightarrow x_i /\lambda$. Hence, 
from any given eigenfunction of ${\cal H}_N$ and ${\cal P}_N$, one can 
generate a one-parameter family of eigenfunctions by scaling all the $x_i$. 
It may also be observed that ${\cal H}_N$ remains 
invariant while ${\cal P}_N$ changes sign if we change the sign of $\eta$ and 
transform all the $x_i \rightarrow - x_i$ at the same time;
such a transformation may be called as `parity transformation'. 
Due to the invariance of ${\cal H}_N$ under this parity transformation, 
it is sufficient to study the eigenvalue problem (\ref{b1}) for
one particular sign of $\eta$, say, $\eta >0$. 
The eigenfunctions for $\eta < 0$ case can then be constructed from 
those for $\eta>0$ case by simply changing $x_i \rightarrow -x_i$;
this leaves all energy eigenvalues invariant but reverses the
sign of the corresponding momentum eigenvalues. 

For the purpose of solving the eigenvalue problem (\ref{b1}) 
through the coordinate Bethe ansatz, it is convenient to divide 
the coordinate space $R^N \equiv \{ x_1, x_2, \cdots x_N \}$ 
into various $N$-dimensional sectors defined through inequalities like
$x_{\omega(1)}< x_{\omega(2)}< \cdots < x_{\omega(N)}$, where 
$\{\omega(1), \omega(2), $ $\cdots , \omega(N)\}$ 
represents a permutation of the integers $\{1,2, \cdots ,N \}$. Since the 
interaction part of the Hamiltonian (\ref{a2}) vanishes within each such 
sector, the resulting eigenfunction can be expressed as a superposition of 
free particle wave functions. The coefficients associated
with these free particle wave functions can be computed by using the
interaction part of the Hamiltonian (\ref{a2}), which is nontrivial only 
at the boundary of two adjacent sectors. It is known that all such
coefficients, which appear in the Bethe ansatz solution of a $N$-particle
system having only local interactions 
(like the $\de$-function or derivative $\de$-function type interactions), 
can be obtained by simply solving the corresponding two-particle
problem \cite{Gu87}. Thus, by using the solutions of the related 
two-particle problem, it is possible to construct completely symmetric
$N$-particle eigenfunctions for the Hamiltonian (\ref{a2}).
In the region $x_1< x_2 < \cdots < x_N$, such eigenfunctions 
can be written in the form \cite{SMB94,Gu87}
\beq \tau_N (x_1, x_2 , \cdots , x_N) ~=~ \sum_\omega \left(\prod_{l<m}
\frac{A(k_{\omega(m)},k_{\omega(l)})}{A(k_m,k_l)}\right) \rho_{\omega(1),
\omega(2), \cdots , \omega(N)} (x_1, x_2, \cdots , x_N) ~, \label{b3} \eeq
where
\beq
\rho_{\omega(1), \omega(2), \cdots , \omega(N)} (x_1, x_2, \cdots , x_N) ~=~
\exp ~\{ i (k_{\omega(1)}x_1 + \cdots + k_{\omega(N)} x_N ) \} ~,
\label{b4} \eeq
$k_n$'s are all distinct quasi-momenta, $\omega$ represents an element of the 
permutation group for the integers $\{ 1,2,....N \}$
and $\sum_{\omega}$ implies summing over all such permutations. 
The coefficient $A(k_l,k_m)$ in Eq.~(\ref{b3}) is obtained 
by solving the two-particle problem related to the derivative $\de$-function
Bose gas and this coefficient is given by
\beq A(k_l,k_m) ~=~ \frac{k_l - k_m + i \eta (k_l+ k_m)}{k_l - k_m} ~.
\label{b5} \eeq
The eigenvalues of
the momentum (\ref{b2}) and Hamiltonian (\ref{a2}) operators, 
corresponding to the eigenfunctions $\tau_N(x_1, x_2, \cdots , x_N)$ of the 
form (\ref{b3}), are easily obtained as 
\bea &&~~~~~~~~{\cal P}_N ~\tau_N(x_1, x_2, \cdots , x_N) ~=~ \hbar
\Big(\sum_{j=1}^N k_j \Big) ~\tau_N(x_1, x_2, \cdots , x_N) ~, \non
~~~~~~~~~~~~~~~~~ (2.6a) \\
&&~~~~~~~~{\cal H}_N ~\tau_N(x_1, x_2, \cdots , x_N) ~=~
\hbar^2 \Big(\sum_{j=1}^N k_j^2 \Big) ~\tau_N(x_1, x_2, \cdots , x_N) ~.
\non ~~~~~~~~~~~~~~~~ (2.6b) \eea
\addtocounter{equation}{1}

It should be noted that, Bethe states of the form (\ref{b3}) represent
scattering as well as bound states for the Hamiltonian (\ref{a2}) of 
the derivative $\de$-function Bose gas. However, for the case of scattering 
states all the $k_j$'s are real numbers, while for the case of bound 
states the $k_j$'s are allowed to take complex values in general. As 
mentioned above, for a translationally invariant system, a wave function 
represents a localized bound state if the corresponding
probability density decays sufficiently fast 
when any of the relative coordinates measuring the distance between
a pair of particles tends towards infinity. To obtain the condition for which 
the Bethe state (\ref{b3}) would represent such a localized bound state, 
let us first consider the following
wave function in the region $x_1<x_2<\cdots <x_N$ :
\bea \rho_{1,2, \cdots ,N} (~ x_1,x_2, \cdots , x_N ~) ~=~ \exp ~(i\sum_{j=1}^N
k_j x_j)\,, \label{b7} \eea
where $k_j$'s in general are complex valued wave numbers.
As before, the momentum eigenvalue corresponding to this wave function is
given by $\hbar\sum_{j=1}^N k_j$. Since this must be a real quantity, one 
obtains the condition
\beq \sum_{j=1}^N q_j ~=~ 0 ~, \label{b8} \eeq
where $q_j$ denotes the imaginary part of $k_j$. By using the 
condition (\ref{b8}), the probability density
corresponding to the wave function $\rho_{1,2, \cdots ,N} (~ x_1,x_2, \cdots
,x_N ~)$ in (\ref{b7}) can be expressed as
\bea {|\rho_{1,2,\cdots ,N} (~ x_1,x_2, \cdots , x_N ~)|}^2 ~=~ \exp ~\Big\{ ~
2 \sum_{r=1}^{N-1} \Big(\sum_{j=1}^r q_j\Big) ~y_r ~\Big \} ~, \label{b9} \eea
where the $y_r$'s are the $N-1$ relative coordinates: $y_r \equiv
x_{r+1} - x_r \, $. It is evident that the
probability density in (\ref{b9}) decays exponentially in the limit
$y_r \rightarrow \infty$ for one or more values of $r$, provided that all
the following conditions are satisfied:
\beq q_1< 0 ~, ~~~~q_1+q_2 < 0 ~, ~~\cdots\cdots ~~, ~\sum_{j=1}^{N-1} ~q_j < 0
~. \label{b10} \eeq
It should be observed that the wave function (\ref{b7}) is obtained by taking 
$\omega$ as the identity permutation in (\ref{b4}). However, the
Bethe state (\ref{b3}) also contains terms like (\ref{b4}) with
$\omega$ representing all possible nontrivial permutations.
The conditions which ensure the decay of such a term, associated
with any nontrivial permutation $\omega$, are evidently given by 
\beq q_{\omega(1)}< 0 ~, ~~~~q_{\omega(1)}+ q_{\omega(2)}< 0 ~, 
~~\cdots\cdots ~~, ~\sum_{j=1}^{N-1} ~q_{\omega(j)} < 0 ~. \label{b11} \eeq
It is easy to check that above conditions, in general, contradict the 
conditions given in Eq.~(\ref{b10}). To bypass this problem and ensure an 
overall decaying wave function (\ref{b3}), it is sufficient 
to assume that the coefficients of all terms
$\rho_{\omega(1), \omega(2), \cdots , \omega(N)}
(x_1, x_2, \cdots , x_N)$ with nontrivial permutations 
take the zero value. This leads to a set of relations given by 
\beq A( k_{r}, k_{r+1} ) ~=~ 0 \, , \quad {\rm for} \quad r \in {\Omega}_N\,,
\label{b12} \eeq
where ${\Omega}_N \equiv \{1,\,2,\,\cdots,\,N-1\}$.
Thus the simultaneous validity of the conditions
(\ref{b8}), (\ref{b10}) and (\ref{b12}) ensures that the Bethe state
$\tau_N(x_1, x_2, \cdots , x_N)$ (\ref{b3}) represents a bound state. 

Let us now discuss how the conditions (\ref{b8}), (\ref{b10}) and (\ref{b12}) 
can be simplified for the case of the derivative $\de$-function Bose gas. 
Using the conditions (\ref{b8}) and (\ref{b12}) along with Eq.~(\ref{b5}), 
one can easily derive an expression for all the quasi-momenta as
\beq k_n ~=~ \chi ~e^{-i(N+1-2n)\phi} ~, \label{b13} \eeq
where $\chi$ is a real, non-zero parameter, 
and $\phi$ is related to the coupling constant $\eta$ as
\beq \phi ~=~ \tan^{-1} (\eta ) ~ \Longrightarrow ~~ \eta = \tan \phi \, .
\label{b14} \eeq
To obtain an unique value of $\phi$ from the above equation, 
it may be restricted 
to the fundamental region $-\frac{\pi}{2} < \phi (\neq 0) < \frac{\pi}{2}$.
Furthermore, since we have seen that ${\cal H}_N$ (\ref{a2}) remains invariant 
under the `parity transformation', it is enough to study the corresponding
eigenvalue problem only within the range $0 < \phi <
\frac{\pi}{2}$. Next, let us consider 
the remaining conditions (\ref{b10}) for the existence of a localized bound 
state. Since summation over the imaginary parts of $k_n$'s in (\ref{b13}) 
yields
\bea \sum_{j=1}^l q_j = -\chi ~\frac{\sin (l \phi)}{\sin \phi} ~\sin [(N-l) 
\phi]\, , \label{b15} \eea
one can rewrite the conditions (\ref{b10}) in the form
\beq \chi ~\frac{\sin (l \phi)}{\sin \phi} ~\sin [(N-l) \phi] ~>~ 0 \, ,
\quad {\rm for} \quad l \in {\Omega}_N\, . \label{b16} \eeq
where ${\Omega}_N$ denotes the set of integers $\{1,\,2,\,\cdots,\,N-1\}$.
Consequently, for any given values of $\phi$ and $N$, a bound state
will exist when all the inequalities in Eq.~(\ref{b16}) 
are simultaneously satisfied for some real non-zero value of $\chi$.

Let us make a comment at this point. Within the region
$x_1< x_2 < \cdots < x_N$, the completely symmetric 
Bethe eigenfunctions associated with the Hamiltonian (\ref{a1}) 
of the $\de$-function Bose gas can also be expressed through Eq.~(\ref{b3}),
where $A(k_l,k_m)$ is given by \cite{LL63,Ta99,Gu87}
\[ A(k_l,k_m) ~=~ \frac{k_l - k_m -i \mu}{k_l - k_m}. \]
Hence, the equations (\ref{b8}), (\ref{b10}) and (\ref{b12}) together give a 
sufficient condition for the existence of
bound states for the case of the $\de$-function Bose gas also. 
By using the above mentioned form of $A(k_l,k_m)$, it is easy to show that 
Eq.~(\ref{b12}) completely fixes the corresponding quasi-momenta as 
equidistant points on a straight line parallel to the imaginary axis in the 
complex momentum plane. Moreover, the condition (\ref{b8}) ensures that such
equidistant points would be symmetric under reflection 
with respect to the real axis in the complex momentum plane. 
It is easy to check that the remaining 
conditions (\ref{b10}) for bound state formation are trivially satisfied 
by those quasi-momenta for any negative value of the coupling constant $\mu$. 
Thus, for any $N \geq 2$, bound states are formed for the case of the 
$\de$-function Bose gas for all negative values of the coupling constant. 
However in the previous paragraph we have seen that, for the case of the
derivative $\de$-function Bose gas, the conditions (\ref{b8}) and (\ref{b12}) 
lead to Eq.~(\ref{b13}) implying that the quasi-momenta associated with a 
bound state are represented through equidistant points on a
circle or circular `string' with an arbitrary radius and having its centre at 
the origin of the complex momentum plane. Consequently, the remaining 
conditions (\ref{b10}) for bound state formation yield Eq.~(\ref{b16}), 
which is quite nontrivial in nature. Indeed, by solving this equation, we have
found earlier that the derivative $\de$-function Bose gas allows bound 
states in only certain non-overlapping ranges of the coupling constant $\eta$
\cite{BBS03,BBS04,BBS04a}. Furthermore,
such non-overlapping ranges of $\eta$ crucially depend on the value of 
$N$ and they can be determined by using the Farey sequences in number theory.

We would now like to find the conditions for constructing clusters of bound 
particles in the case of the derivative $\de$-function Bose gas. To this end, 
we shall first discuss the concept of a `clustered state' for 
any translationally invariant system, and then give a prescription
for finding the Bethe states representing clusters of bound particles. 
Let us consider a system of $N$ particles which are divided
into some groups or clusters --- with at least one group containing more than
one particle. It is assumed that particles within the same group
behave like the constituents of a bound state, but particles corresponding 
to different groups behave like the constituents of a scattering state.
More precisely, a wave function corresponding to such an $N$-particle system 
satisfies the following two conditions. If the relative distance between
any two particles belonging to the same group goes to infinity, the
probability density corresponding to the $N$-particle wave function
decays in the same way as a bound state. 
On the other hand, if the relative distance between any two particles
belonging to different groups tends towards infinity, the probability
density remains finite similar to a scattering state.
If any wave function corresponding to a $N$-particle system 
satisfies these two conditions, we define it as a clustered state.
It is interesting to observe that, 
for the case of the $\de$-function Bose gas, the Bethe states corresponding
to clusters of bound particles can always be expressed as linear
superpositions of several clustered and bound states, with the restriction 
that at least one clustered state must be present in such a superposition
\cite{Ta99,Do10}. This observation may be used to define the Bethe states 
corresponding to clusters of bound particles for the case 
of any translationally invariant system. More precisely, we assume that
a Bethe state for any translationally invariant system would represent
clusters of bound particles if it can be expressed as some linear 
superposition of clustered and bound states, with the restriction that at 
least one clustered state must be present in this superposition.

Next, let us discuss how the conditions for constructing a clustered state
can be implemented for the case of the plane wave function (\ref{b7}). Since 
any eigenvalue of the momentum operator ${\cal P}_N$ must be a real quantity, 
Eq.~(\ref{b8}) is also obeyed for this case. To proceed further,
let us choose a specific value of $N$ given by $N=4$. 
For this case, the probability density (\ref{b9}) may be explicitly written as 
\beq {|\rho_{1,2,\cdots ,4} (~ x_1,x_2, \cdots , x_4 ~)|}^2 ~=~ \exp~\{ 
2q_1y_1 + 2(q_1+q_2)y_2 + 2(q_1+q_2+q_3)y_3 \} ~. \label{b17} \eeq
Suppose, the conditions (\ref{b10}) for a bound state formation 
are slightly modified for this case as 
\[ q_1<0, ~~~~q_1+q_2 =0, ~~~~q_1+q_2+q_3 <0 \, . \]
Taking into account this new condition, it is easy to see that when
$y_1 = x_2 - x_1$ or $y_3 = x_4 - x_3$ tends towards infinity, the probability 
density in Eq.~(\ref{b17}) still decays like a bound state. 
On the other hand, when $y_2 = x_3-x_2$ tends towards
infinity, the probability density in Eq.~(\ref{b17}) remains finite. Hence,
the clusters of particles given by $\{1,2\}$ and $\{3,4\}$ satisfy all 
the criteria of a clustered state. Generalizing this specific example, we 
replace some of the inequalities in Eq.~(\ref{b10}) by equalities. 
More precisely, we modify the conditions in Eq.~(\ref{b10}) as 
\bea &&~~~~~~~~~~~~~~~~~~~~~~~~\sum_{i=1}^l q_i = 0 \, , ~~{\rm
for}~ l~ \in {\tilde \Omega}_{N} \, ,~~~~~~~~~~~~~~~~~~~~~~~~~~~~~~~~~~~~~
(2.18a) \non \\
&& ~~~~~~~~~~~~~~~~~~~~~~~~\sum_{i=1}^l q_i ~<~ 0 \, , ~~ {\rm for}
~l~\in (\Omega_N - {\tilde \Omega}_N)\, ,
~~~~~~~~~~~~~~~~~~~~~~~~~(2.18b)\non \eea
\addtocounter{equation}{1}
where $\tilde \Omega_{N}$ is any non-empty proper subset of $\Omega_N$ and
$(\Omega_N - \tilde \Omega_{N})$ is the complementary set of 
$\tilde \Omega_{N}$. Let us now assume that the quasi-momenta 
associated with the plane wave function (\ref{b7})
satisfy the relations (\ref{b8}) and (2.18a,b), where the set $\tilde 
\Omega_{N}$ contains $p$ number of elements. Due to Eq.~(\ref{b9}) it is 
evident that this plane wave function would represent 
a clustered state containing $(p+1)$ number of clusters. 

Finally, we try to find the simplest possible condition for which 
the Bethe state (\ref{b3}) would represent clusters of bound particles.
Let us assume that the quasi-momenta corresponding to this Bethe state 
satisfy the relations (\ref{b12}). As a result, the coefficients of all plane 
waves except (\ref{b7}) take the zero value within the Bethe state (\ref{b3}).
Thus, due to the relations (\ref{b12}), 
the Bethe state (\ref{b3}) would reduce to the plane wave function (\ref{b7}). 
Next, we assume that the quasi-momenta corresponding to this 
Bethe state also satisfy the relations (\ref{b8}) and (2.18a,b).
Hence, the plane wave function (\ref{b7}) represents a clustered state.
Consequently, Eqs.~(\ref{b8}), (\ref{b12}) and (2.18a,b) together 
yield a sufficient condition for which the Bethe state (\ref{b3}) 
would represent clusters of bound particles. It is obvious that such a 
Bethe state is expressed through a single clustered state, instead of a 
linear superposition of several clustered and bound states. 

We would like to make a remark at this point. It may be noted that, 
for the case of the $\de$-function Bose gas, it is not possible to find any 
set of quasi-momenta which simultaneously satisfy the equations
(\ref{b8}), (\ref{b12}) and (2.18a,b). Indeed we have already mentioned
that, for this case, the conditions (\ref{b8}) and (\ref{b12}) completely fix
the corresponding quasi-momenta as equidistant and reflection symmetric points 
on a straight line parallel to 
the imaginary axis in the complex momentum plane. It is easy to check 
that such quasi-momenta do not satisfy Eq.~(2.18a). To bypass this problem, it 
should be noted that Eqs.~(\ref{b8}), (\ref{b12}) and (2.18a,b) together 
yield only a {\it sufficient} condition, but not the {\it necessary} condition
for which a Bethe state leads to clusters of bound particles. In fact, 
it is possible to take a partially relaxed version of Eq.~(\ref{b12}) given by
\beq A( k_{r}, k_{r+1} ) ~=~ 0 \, , \quad {\rm for} \quad r \in {\Omega}'_N\,,
\label{b19} \eeq
where ${\Omega}'_N$ is any non-empty proper subset of ${\Omega}_N$.
However, the corresponding Bethe state (\ref{b3}) not only contains the plane 
wave (\ref{b7}), but also plane waves like (\ref{b4}) with
$\omega$ taking values within a subset of all possible 
nontrivial permutations. The Bethe state (\ref{b3}) would represent
clusters of bound particles if each of these plane waves associated 
with nontrivial permutations behaves like either a bound state or a clustered 
state. Hence, as a consequence of taking Eq.~(\ref{b19}) instead of 
Eq.~(\ref{b12}), Eq.~(\ref{b11}) or analogues of Eqs.~(2.18a,b) for the 
required values of nontrivial permutation $\omega$ must also be satisfied.
For the case of the $\de$-function Bose gas, a 
solution of Eqs.~(\ref{b8}) and (\ref{b19}) yields $N$ number of quasi-momenta 
which can be represented through reflection symmetric  
points on several (more than one) `strings' parallel to the imaginary axis in 
the complex momentum plane. It can be shown that, apart from satisfying 
Eqs.~(2.18a,b), those quasi-momenta also satisfy Eq.~(\ref{b11}) or analogues 
of Eqs.~(2.18a,b) for the required values of nontrivial permutation $\omega$. 
As a result, the Bethe state corresponding to those quasi-momenta
represents clusters of bound particles for the case of the $\de$-function Bose 
gas. Evidently, such a Bethe state is expressed through 
a superposition of several clustered and bound states. 

In the next section our aim will be to show that, unlike the 
case of the $\de$-function Bose gas, it is possible to simultaneously solve 
the equations (\ref{b8}), (\ref{b12}) and (2.18a,b) for the case of 
the derivative $\de$-function Bose gas. 
To this end, let us discuss how to simplify these three equations 
for the latter case. Since the exact form of the set $\tilde \Omega_{N}$ 
appearing in (2.18a,b) may depend on the value of $\phi$ for the case of the 
derivative $\de$-function Bose gas, in the following we shall replace the 
notation ${\tilde \Omega}_N$ by $ \Omega_{N, \phi}$.
Using Eqs.~(\ref{b8}) and (\ref{b12}) along with the form of $A(k_l,k_m)$ 
given in (\ref{b5}) it is easy to see that, similar to the case of a 
localized bound state, the quasi-momenta associated with
clusters of bound particles can be written in the form (\ref{b13}) and the 
imaginary parts of these quasi-momenta satisfy the relation (\ref{b15}).
By using Eq.~(\ref{b15}), we can recast the remaining 
conditions (2.18a,b) for the formation of
clusters of bound particles (for any given values of $\phi$ and $N$) as
\bea &&~~~~~\chi ~\frac{\sin (l \phi)}{\sin \phi} ~\sin [(N-l) \phi] = 0 
\, ,\quad {\rm
for}~~~ l~ \in {\Omega}_{N, \phi} \, ,~~~~~~~~~~~~~~~~~~~~~~~~~~~~~~~~
(2.20a) \non \\
&&~~~~~\chi ~\frac{\sin (l \phi)}{\sin \phi} ~\sin [(N-l) \phi] ~>~ 0 \, ,
\quad {\rm for} ~~~l~\in (\Omega_N - \Omega_{N,\phi})\, ,
~~~~~~~~~~~~~~~~~~~~(2.20b)\non \eea
\addtocounter{equation}{1}
where $\chi$ is any non-zero real number, 
$\Omega_{N, \phi}$ is any non-empty proper subset of $\Omega_N$ and
$(\Omega_N - \Omega_{N, \phi})$ is the complementary set of 
$\Omega_{N, \phi}$. Let us assume that Eqs.~(2.20a,b) are satisfied
for some values of $\phi$ and $N$, where $\Omega_{N, \phi}$ is given by 
\beq \Omega_{N, \phi} = \{ l_1,l_2, \cdots , l_p \} \, , \label{b21} \eeq
with $1 \leq p < N-1$.
Then from Eq.~(\ref{b9}) it follows that the sets of particles given by
$\{ 1, \cdots ,l_1 \}, \{ l_1+1 , \cdots ,l_2 \}, \cdots \cdots ,
\{ l_{p-1}+1 , \cdots ,l_p \}, \{ l_{p}+1, \cdots , N \}$
represent $(p+1)$ number of clusters of bound particles.
Moreover, the numbers of particles present within each of these clusters,
i.e., the size of the clusters, may be written in the form
\beq \{\!\{ \, l_1, l_2-l_1, \cdots , l_p-l_{p-1}, N- l_p \, \}\!\} \, .
 \label{b22} \eeq

Since the quasi-momenta associated with both bound states and 
clusters of bound particles are given by Eq.~(\ref{b13}), the momentum and 
energy eigenvalues for clusters of bound particles can be derived in exactly 
the same way as has been done earlier \cite{BBS03} for the case of a bound 
state. Inserting the quasi-momenta given in Eq.~(\ref{b13}) to Eqs.~(2.6a,b),
we obtain the momentum eigenvalue as
\beq P ~=~ \hbar \chi ~\frac{\sin (N\phi)}{\sin \phi} ~, \label{b23} \eeq
and the energy eigenvalue as
\beq E ~=~ \frac{\hbar^2 \chi^2 \sin(2N \phi)}{\sin(2\phi)} ~. \label{b24} \eeq

\noi \section{Farey sequences and clusters of bound particles}
\renewcommand{\theequation}{3.{\arabic{equation}}}
\setcounter{equation}{0}

\medskip

In this section, we shall try to find all possible solutions of the sufficient
conditions (2.20a,b) for constructing clusters of bound particles
in the case of the derivative $\de$-function Bose gas. Some properties
of the Farey sequences \cite{ZZM00} in number theory will play a crucial role 
in our analysis. Due to the existence of the parity transformation,
as mentioned in the earlier section, it is sufficient to concentrate on values
of $\phi$ lying in the range $0<\phi <\frac{\pi}{2}$. 
Within this range of $\phi$, $\sin\phi>0$ and hence the conditions (2.20a,b)
for forming clusters of bound particles reduce to 
\bea &&~~~~\chi ~\sin (l \phi)~\sin [(N-l) \phi] = 0 \, , \quad \quad {\rm
~~for}~~~l~\in \Omega_{N, \phi} \, ,~~~~~~~~~~~~~~~~~~~~~~~~~~~~~
~~(3.1a) \non \\
&&~~~~\chi ~\sin (l \phi)~\sin [(N-l) \phi] ~>~ 0 \, ,\quad\quad {\rm
for} ~~l ~ \in (\Omega_N - \Omega_{N, \phi})\,. 
\, ~~~~~~~~~~~~~~~~~~~~~~(3.1b)\non \eea
\addtocounter{equation}{1}
Let us first try to find the values of $\phi$ 
for which Eq.~(3.1a) holds true. For any given $\phi$,
this equation would be satisfied for some value of $l$ if either 
$\sin l\phi = 0$ or $\sin [(N-l)\phi] =0$.
For $\sin l\phi=0$, $l\phi = k\pi$ and so $\phi/\pi = k/l$ where $k$ is an
integer. Since $l \in \Omega_{N,\phi}$, the denominator in the expression of
$\phi/\pi ~(=k/l)$ is always less than $N$. Similarly, for $\sin
[(N-l)\phi]=0$, $\phi/\pi = m/(N-l)$ where $m$ is an integer. Since $l \in
\Omega_{N, \phi}$, in this case also, the denominator in the expression of 
$\phi/\pi ~(= m/(N-l))$ is less than $N$ for all
values of $l$. Hence it follows that, the condition (3.1a) would be
satisfied if and only if $\phi/\pi$ can be expressed in the form 
\bea \frac{\phi}{\pi} = \frac{a}{b}\,, \label{c2} \eea
where $\{a,b\}$ are relatively prime integers (i.e, the greatest common
divisor of $a$ and $b$ is 1), taking values within the ranges 
\bea ~~~~~~~~~~~~~~~~~~~~~~~~~~~~~~~0< a < \frac{b}{2}\,, ~~~~2 < b \leq N-1\,.
~~~~~~~~~~~~~~~~~~~~~~~~~~~~~~~~(3.3a,b) \non 
\eea
\addtocounter{equation}{1}
Due to Eq.~(3.3b) it is evident that, clusters of bound particles 
can exist only for $N \geq 4$. In the following, we shall establish a 
connection of the fractions $\phi/\pi$, given by Eqs.~(\ref{c2}) and (3.3a,b),
with the Farey sequences in number theory. 

For a positive integer $N$, the Farey sequence is defined to be the set of all
the fractions $a/b$ in increasing order such that (i) $0 \le a \le b \le N$,
and (ii) \{$a, b$\} are relatively prime integers \cite{ZZM00}. 
The Farey sequences for the first few values of $N$ are given by
\bea F_1: & & \quad \frac{0}{1} ~~~~\frac{1}{1} \non \\
F_2: & & \quad \frac{0}{1} ~~~~\frac{1}{2} ~~~~\frac{1}{1} \non \\
F_3: & & \quad \frac{0}{1} ~~~~\frac{1}{3} ~~~~\frac{1}{2} ~~~~\frac{2}{3}
~~~~
\frac{1}{1} \non \\
F_4: & & \quad \frac{0}{1} ~~~~\frac{1}{4} ~~~~\frac{1}{3} ~~~~\frac{1}{2}
~~~~
\frac{2}{3} ~~~~\frac{3}{4} ~~~~\frac{1}{1} \non \\
F_5: & & \quad \frac{0}{1} ~~~~\frac{1}{5} ~~~~\frac{1}{4} ~~~~\frac{1}{3}
~~~~
\frac{2}{5} ~~~~\frac{1}{2} ~~~~\frac{3}{5} ~~~~\frac{2}{3} ~~~~
\frac{3}{4} ~~~~\frac{4}{5} ~~~~\frac{1}{1}
\label{c4} \eea
These sequences enjoy several properties, of which we list the relevant ones
below. \\
\noi (i) Let $a/b, a'/b'$ are two fractions appearing in the Farey
sequence $ F_N$. Then $a/b < a'/b' ~(~ a'/b' < a/b~) $ are two successive 
fractions in $F_N$, if and only if the following two conditions are satisfied:
\bea &&~~~~~~~~~~~~~~~~~~~~~~~~~~~~~~~~~a' b ~-~ a b' ~=~ 1~ (-1)~, ~~~~
~~~~~~~~~~~~~~~~~~~~~~~~~~~~~~~~~~(3.5a) \non \\
&&~~~~~~~~~~~~~~~~~~~~~~~~~~~~~~~~~b+b'~>~N\,.
 ~~~~~~~~~~~~~~~~~~~~~~~~~~~~~~~~~~~~~~~~~~~~~~~~~~~
(3.5b) \non \eea
\addtocounter{equation}{1}
It then follows that both $a$ and $b'$ are relatively prime to $a'$ and $b$.

\noi (ii) For $N \ge 2$, if $n/N$ is a fraction appearing somewhere in the
sequence $F_N$ (this implies that $N$ and $n$ are relatively prime according to
the definition of $F_N$), then the fractions $a_1/b_1$ and $a_2/b_2$ appearing
immediately to the left and to the right respectively of $n/N$ satisfy
\bea a_1 ~,~ a_2 ~\le ~n ~, \quad {\rm and} \quad a_1 ~+~ a_2 ~=~ n ~, \non \\
b_1 ~,~ b_2 ~<~ N ~, \quad {\rm and} \quad b_1 ~+~ b_2 ~=~ N ~. \label{c6} \eea

To apply the above mentioned Farey sequence in the present context, 
let us define a subset of $F_N$ as 
\bea {F}^\prime_N = \left \{ \left.\frac{a}{b} ~\right| ~~\frac{a}{b} \in F_N,
 ~~~\frac{0}{1}<\frac{a}{b} < \frac{1}{2} \right\}\,, \label{c7} \eea
and a subset of $F^\prime_N$ as 
\bea F''_N = \left\{ \left.\frac{n}{N} ~\right| ~~ \frac{n}{N} \in F^\prime_N 
\right \}\,. \label{c8} \eea
Using these definitions of various subsets of a Farey sequence, we find that
\bea F^\prime_N = F^\prime_{N-1} \cup F''_N\,. \label{c9} \eea
Furthermore, it is worth noting that,
 Eqs.~(\ref{c2}) and (3.3a,b) can equivalently be expressed as 
\bea \frac{\phi}{\pi} \in F^\prime_{N-1} \,. \label{c10} \eea
Consequently, it follows that the condition (3.1a) for cluster
formation is obeyed if and only if $\phi/\pi \in F'_{N-1}$.
In this context it may be observed that, due to Eq.~(\ref{c9}),
all the elements of $F^\prime_{N-1}$ are also present in $F^\prime_N$. 
By using such an embedding of $F^\prime_{N-1}$ into $F^\prime_N$,
we find that any fraction $a/b \in F'_{N-1}$ belongs to one of the
four distinct classes, which are defined in the following:
\vskip .11 cm 

\noi 
I. At least one of the fractions nearest to $a/b$ (from either the left or 
the right side) in the sequence $F^\prime_N$ lies in the set $F''_N$.
Then, from a property of the Farey sequences, it follows that
$\{b, N\}$ are relatively prime integers in this case.
\\
II. None of the nearest fractions of $a/b$ (from the left or right
side) in the sequence $F'_N$ lies in the 
set $F''_N$, and $\{b, N\}$ are relatively prime integers. 
\\
III. $N$ is divisible by $b$. Clearly, $\{b, N\}$ are not relatively prime
integers in this case.
\\
IV. $N$ is not divisible by $b$, and $\{b,N\}$ are not relatively prime
integers.
\vskip .11 cm 

\noi 
To demonstrate the above mentioned classification through an 
example, let us choose $N=6$. For this case, the sets
$F^{\prime}_5$, $F^{\prime}_6$ and $F''_6$ are given by
\bea F^\prime_5 :& &\frac{1}{5}~~~~\frac{1}{4}~~~~\frac{1}{3}~~~~
\frac{2}{5} \non \\
F^\prime_6 :& &\frac{1}{6}~~~~\frac{1}{5}~~~~\frac{1}{4}~~~~~
\frac{1}{3}~~~~\frac{2}{5}\,; ~~~~~F''_6 :~~ \frac{1}{6}\,.\non \eea
Using the embedding of $F^{\prime}_5$ into $F^{\prime}_6$,
it is easy to verify that each fraction in $F^\prime_5$ falls under one
of the four classes discussed above. More precisely, the fractions 1/5,
1/4, 1/3 and 2/5 belong to type I, type IV, type III and type II respectively. 
Returning back to the general case we note that, for any fraction $a/b \in 
F'_{N-1} \,$, $\{b,N\}$ are either relatively prime integers 
or not relatively prime integers. 
If $\{b,N\}$ are relatively prime integers, then it is obvious that
$a/b$ must be an element of either type I or type II. On the other hand, 
if $\{b,N\}$ are not relatively prime integers, then 
$a/b$ must be an element of either type III or type IV. In this way,
one can show that any fraction $a/b \in F'_{N-1}$ belongs to one of these
four distinct classes. This type of classification 
for the elements of $F^\prime_{N-1}$ will shortly play an important role 
in our analysis on the formation of clusters of bound particles. 

Next, we try to find the elements of $F'_{N-1}$ which would satisfy the 
remaining condition (3.1b) for cluster formation. For any $a/b \in 
F^{\prime}_{N-1}$, one can express the rational number $Na/b$ as 
\bea \frac{Na}{b} = \left[ \frac{Na}{b}\right ] + \frac{\rho}{b} \,,
\label{c11} \eea
where $[x]$ denotes the integer part of $x$ and $\rho \in \{0, 1, \cdots
b-1\}$. The above equation can also be written in the form
\beq aN - bt = \de\,, \label{c12} \eeq
where the integers $t$ and $\de$ are defined as 
\bea
&&t = \left [\frac{Na}{b}\right],~\de = \rho\,, {\, ~~~~~~~~~~~~~~~~\rm if}
~~~\frac{\rho}{b} \leq \frac{1}{2}\,, \non \\
&&t = \left [\frac{Na}{b}\right] + 1, ~\de = \rho - b\,, {~~~~~~~\rm if}
~~~~\frac{\rho}{b} > \frac{1}{2}\,.
\label{c13} \eea
One can derive several bounds on the values of $t$ and $\de$.
By using Eq.~(\ref{c13}), it is easy to show that
\beq \frac{|\de|}{b} \leq \frac{1}{2}\, . \label{c14} \eeq
Using Eqs.~(3.3a,b) we obtain $Na/b\geq N/(N-1)>1$, which in turn yields 
$t\geq 1$. Moreover, with the help of Eqs.~(\ref{c12}) and (\ref{c13}), we 
also find that 
\bea \frac{t}{N} < \frac{1}{2}\,. \label{c15} \eea
The derivation of Eq.~(\ref{c15}) is given in Appendix A.

It may be observed that, for any given $a/b \in F^{\prime}_{N-1}$, the values 
of $t$ and $\de$ are uniquely determined through Eqs.~(\ref{c11}) and 
(\ref{c13}). Thus, we get a mapping of the form 
\bea (a,b,N) \longrightarrow (t, \de)\,. \label{c16} \eea
In the following, we shall show that this mapping considerably simplifies
the analysis of Eq.~(3.1b) by casting it in an alternative form.
To this end, let us define a function $f(l, N, \phi)$ as
\bea f(l, N, \phi)= \chi \, \sin l\phi \, \sin ( N-l )\phi \,. \label{c17} \eea
Substituting the expression of $\phi$ given in Eq.~(\ref{c2})
to the above equation, we get 
\bea f(l, N, \phi) = \chi \sin \left(\frac{\pi a l}{b} \right) \sin \left \{
\frac{\pi a}{b}( N-l )\right\}\,. \non \eea
Next, by using the relation (\ref{c12}), we express $f(l, N, \phi)$ in the form
\bea f(l, N, \phi) = \chi {(-1)}^{t+1} \sin \left( \frac{\pi a l}{b} \right) 
\sin \left \{ \frac{\pi a}{b}\left(l - \frac{\de}{a} \right) \right \} \,. 
\label{c18} \eea 
Let us now define a function $g(x, \phi)$ as
\bea g(x, \phi) = \sin \left( \frac{\pi a x}{b} \right)\,, \label{c19} \eea
where $x$ denotes a continuous real variable. Then Eq.~(\ref{c18}) can be
written as 
\bea f(l, N, \phi) = \chi {(-1)}^{t+1}g(l, \phi) g(l- \bigtriangleup, \phi)\,,
\label{c20} \eea
where the parameter $\bigtriangleup$ is given by
\bea \bigtriangleup = \frac{\de}{a} \,. \label{c21} \eea 
Note that the zero values of the function $g(x, \phi)$ in the variable
$x$ are given by
\bea
x = \frac{mb}{a}, ~~~~~~~~~m = 0, \pm 1, \pm 2, \cdots \,. \label{c22} \eea
Let us consider a value of $l$ such that $f(l, N, \phi) \neq 0$, i.e, 
$l \in \Omega_N - \Omega_{N, \phi}$. Due to Eq.~(\ref{c20}), it is evident 
that both $g(l, \phi)$
and $g(l-\bigtriangleup, \phi)$ take nonzero values for such $l$. 
Since $g(x, \phi)$ is a continuous function of the variable $x$, we can write
\bea sgn \left [ g(l - \bigtriangleup, \phi) \right ] = {(-1)}^{\rho(l)}sgn
\left [ g(l, \phi) \right ] \, , \label{c23} \eea
where {\it sgn} denotes the sign function and 
$\rho(l)$ represents the number of zero points of the function $g(x,\phi)$
within the interval $l-\bigtriangleup < x < l $ 
($ l < x < l - \bigtriangleup $) when $\bigtriangleup > 0$ 
($\bigtriangleup <0$). It is evident that $\rho(l)=0$ for all values of $l 
\in \Omega_N - \Omega_{N, \phi}$, when $\bigtriangleup = 0$. 
Now, by using Eqs.~(\ref{c20}) and (\ref{c23}), we find that
\bea sgn \left [ f(l, N, \phi) \right ] = sgn
\left [ \chi {(-1)}^{t+1+\rho(l)}\right ] \, . \label{c24} \eea
With the help of the above equation, 
we can express Eq.~(3.1b) in an alternative form given by
\bea sgn \left [ \chi {(-1)}^{t+1+\rho(l)} \right ] = 1 \,, 
 ~~\text{~for all~} l \in \Omega_N - \Omega_{N, \phi} \, . \label{c25} \eea
Due to Eq.~(\ref{c22}), it follows that the distance between two consecutive 
zero points of the function $g(x, \phi)$ is given by $b/a$. Moreover, by 
using Eqs.~(\ref{c14}) and (\ref{c21}), we find that 
\bea \frac{b}{a} \geq 2 |\bigtriangleup|\,. \label{c26} \eea
Hence, the value of $\rho(l)$ can be either 0 or 1 for each $l$. 
Since the parameter $\chi$ does not depend on the value of $l$, it is evident
that Eq.~(\ref{c25}) would be satisfied if either 
\bea \rho(l) = 0\,, ~\text{~for all~} l \in \Omega_N - \Omega_{N, \phi} \, , 
\label{c27} \eea
and $\chi$ is chosen such that $sgn (\chi) = {(-1)}^{t+1}$, or 
\bea \rho(l) = 1\,, ~\text{~for all~} l \in \Omega_N - \Omega_{N, \phi} \, ,
\label{c28} \eea
and $\chi$ is chosen such that $sgn (\chi) = {(-1)}^t $. 
However, for any $a/b \in F'_{N-1}$, it can be shown that there exists
at least one $l \in \Omega_N - \Omega_{N, \phi}$ for which $\rho(l) = 0$. The
proof of this statement is given in Appendix B. Thus, it is never possible to 
satisfy the condition given in Eq.~(\ref{c28}). Consequently, for any given 
$a/b \in F'_{N-1}$, clusters of bound particles can be obtained if and only if
Eq.~(\ref{c27}) is satisfied and $\chi$ is chosen such that $sgn (\chi) = 
{(-1)}^{t+1}$. On the other hand, clusters of bound particles cannot be formed
if Eq.~(\ref{c27}) is violated, i.e., if the following condition is satisfied:
\bea \rho(l) = 1\,, ~\text{~for at least one~} l 
\in \Omega_N - \Omega_{N, \phi} \, . \label{c29} \eea 

Previously, we have shown that all elements of $F^{\prime}_{N-1}$ can be
divided into four distinct classes. In the following, we shall analyze each 
class separately and examine whether the elements belonging to each class 
satisfy Eq.~(\ref{c27}) or Eq.~(\ref{c29}).

\vspace {0.5 cm}
\noi {\bf Analysis of Case I}:~~ Let $\phi/ \pi = a/b$ 
be a fraction of type I within the set $F^{\prime}_{N-1}$. In
this case, there exists a fraction $n/N \in F''_N $, such that 
\bea F'_N :
~\cdots \cdots \frac{n}{N} ~~ \frac{a}{b} \cdots \cdots \,{\rm or} 
~ \cdots \cdots \frac{a}{b}~~\frac{n}{N} \cdots \cdots~~ . \non \eea
Hence, using the property of Farey sequences given in Eq.~(3.5a), we get 
\bea aN-bn = \pm 1\,. \label{c30} \eea
Comparing the above equation with (\ref{c12}), we find that 
\bea \frac{\de}{b} = n-t\pm \frac{1}{b}\,. \label{c31} \eea
Combining Eqs.~(\ref{c31}) and (\ref{c14}), one obtains an 
inequality of the form 
\bea \left | n-t \pm \frac{1}{b} \right | \leq \frac{1}{2}\,. \label{c32} \eea
Since, due to Eq.~(3.3b) it follows that $1/b \leq 1/3$, 
one can easily show that Eq.~(\ref{c32}) would be 
satisfied if and only if $t=n$. Since $\{n,N\}$ are relatively prime integers,
it follows that $\{t,N\}$ are relatively prime integers for all fractions
of type I. Substituting $n$ in the place of $t$ in Eq.~(\ref{c12}), we get
\bea aN - bn = \de \,. \label{c33} \eea
Comparing Eqs.~(\ref{c30}) and (\ref{c33}), we find that for any fraction of
type I, the value of $\de$ is given by
\bea \de = \pm 1 \,. \label{c34} \eea

Next, for the sake of convenience, we consider only those fractions of type I,
which yield $\de = +1$. By using Eq.~(\ref{c21}), we obtain the corresponding
value of $\bigtriangleup$ as
\bea \bigtriangleup = \frac{1}{a}\,. \label{c35} \eea
Let us take any value of $l$ such that $f(l, N,
\phi) \neq 0$, i.e., $l \in \Omega_N - \Omega_{N, \phi}$. Using 
Eqs.~(\ref{c20}), (\ref{c22}) and (\ref{c35}), we find
that this $l$ satisfies the relations 
\bea l \neq \frac{mb}{a}\,,~~~~~~~l-\frac{1}{a} \neq \frac{mb}{a}\,,
\label{c36} \eea
for any integer value of $m$. In the following, we shall try to find 
if any zero points of the function $g(x, \phi)$ exists within
the range $l-1/a < x < l$, i.e., whether $mb/a \in (l-1/a, l)$
for any integer value of $m$. 
To this end, 
we express $mb/a$ in the form 
\bea \frac{mb}{a} = \left [ \frac{mb}{a} \right ] + \frac{c}{a}\,,
\label{c37} \eea
where 
$c \in \{0, 1,2, \cdots, a-1 \}$. Let us now assume that $mb/a \in (l-1/a,
l)$. Then, by using Eq.~(\ref{c37}), we get the relation 
\bea l - \frac{1+c}{a}< \left [ \frac{mb}{a} \right ] < l - \frac{c}{a}\, .
\label{c38} \eea

However, for $c \in \{0, 1,2, \cdots, a-1 \}$, one also finds that 
\[ l-1\leq l - \frac{1+c}{a}, ~~~~ l -\frac{c}{a} \leq l \, . \]
Combining the above inequalities with those given in Eq.~(\ref{c38}), 
we obtain the relation
\[ l - 1< \left [ \frac{mb}{a} \right ] < l \, , \]
which evidently leads to a contradiction. Hence it is established that, for 
any fraction of type I which gives $\de = 1$ and 
for any $l \in \Omega_N - \Omega_{N, \phi}$, the condition $mb/a \in
(l-1/a, l)$ can never be satisfied. 
As a result, Eq.~(\ref{c27}) is obeyed for all fractions of type I
which yield $\de = 1$. Similarly, it can be shown that Eq.~(\ref{c27}) is 
also obeyed for all fractions of type I which yield $\de = -1$. 
Consequently, we find that clusters of bound particles
are formed for all fractions of type I. We have already 
seen that the relation $t=n$ holds for all of these fractions. Therefore,
according to the discussion just below Eq.~(\ref{c27}), 
the sign of $\chi$ should be chosen for this type of 
clusters of bound particles as $sgn (\chi) = {(-1)}^{n+1}$.

\noi {\bf Analysis of Case II}:~~Let us now consider the fractions of
type II within the set $F^{\prime}_{N-1}$. At first our aim is to show that 
\bea |\de| > 1\,, \label{c39} \eea
for all fractions of this type. By using Eqs.~(\ref{c11}) and (\ref{c13}), 
one can calculate the values of $t$ and $\de$ for these fractions.
In contrast to the case of fractions of type I for which $\{t,N\}$ 
are always relatively prime integers, for the present case 
$\{t,N\}$ may or may not be relatively prime integers. As an example, let us 
take $N=7$, for which $a/b= 1/5$ is a type II fraction.
Since $7/5=1 + 2/5$, using (\ref{c13}) we find that $t=1$ and 
$\de=2$. Hence, $\{t,N\}$ are relatively prime
integers in this case. On the other hand, we may also take $N=6$, 
for which $a/b= 2/5$ is a type II fraction.
Since $6\times 2/5=2 + 2/5$, it follows that $t=2$ and 
$\de=2$. Hence, $\{t,N\}$ are not relatively prime integers in this case. 

Let us first consider all fractions of type II within the set 
$F^{\prime}_{N-1}$, for which $\{t,N\}$ are relatively prime
integers. As we have seen that $t\geq 1$ 
and Eq.~(\ref{c15}) also gives $t/N <1/2$, it is evident that
\bea \frac{t}{N} \in F''_N \subset F'_N\,. \label{c40} \eea
However, since $a/b$ is a fraction of type II, $a/b$ and $t/N$ cannot be two
consecutive fractions in the sequence $F'_N$. Therefore, at least one of the
two relations given in Eqs.~(3.5a) and (3.5b) must be violated if we choose
$a'=t,~ b'=N$. Since Eq.~(3.5b) is obviously satisfied for this case,
Eq.~(3.5a) must be violated. As a result, we obtain 
\bea |aN - bt| \neq 1\,. \label{c41} \eea
Next, let us assume that $aN-bt = 0$. Since $\{a,b\}$ and $\{t,N\}$ are
both relatively prime pairs, the relation $aN-bt = 0$ 
implies that $a=t$ and $b=N$.
However, since $a/b \in F'_{N-1}$, the relation $b=N$ evidently leads to a
contradiction. Thus, it follows that 
\bea aN-bt \neq 0\,. \label{c42} \eea
Since $a, b, t, N$ are all integer numbers, combining Eqs.~(\ref{c41}) and
(\ref{c42}) we find that 
\bea |aN -bt| > 1\,. \label{c43} \eea
Using Eqs.~(\ref{c43}) and (\ref{c12}), one can easily establish the validity
of Eq.~(\ref{c39}).

Next, we consider all fractions of type II within the set $F^{\prime}_{N-1}$, 
for which $\{t,N\}$ are not relatively prime integers.
In this case, one can express $t$ and $N$ as 
\bea ~~~~~~~~~~~~~~~~~~~~~~~~~~~~~~~~~~~~~~~~~~~~t = \alpha t' \, , ~~ N = 
\alpha N' \, ,~~~~~~~~~~~~~~~~~~~~~~~~~~~~~~(3.44a,b) \non \eea
\addtocounter{equation}{1}
where $\{t', N'\}$ are relatively prime integers and $\alpha ~(>1)$ 
is an integer. Using Eqs.~(\ref{c12}) and (3.44a,b), we obtain 
\bea |\de | \, = \, \alpha \, |aN' - bt'|\,. \label{c45} \eea
Let us now assume that $aN' - bt'= 0$, i.e, $a/b = t'/N'$. Since $\{a,b\}$
and $\{t',N'\}$ are both relatively prime pairs, the relation $a/b = t'/N'$
implies that $a=t'$ and $b=N'$. Thus, by using Eq.~(3.44b) we obtain $N =
\alpha b$. However, the relation $N=\alpha b$ clearly contradicts our
initial assumption of $a/b$ being a fraction of type II, for which $\{b, N\}$
must be relatively prime integers. Consequently, we find that $aN' - bt'
\neq 0$. Since $a, b, t', N'$ are all integer numbers, it follows that 
\bea |aN' - bt'| \, \geq \, 1 \,. \label{c46} \eea
Using Eqs.~(\ref{c45}), (\ref{c46}) and the relation $\alpha>1$, 
one can easily establish the inequality given in (\ref{c39}).

Next, our aim is to show how Eq.~(\ref{c29}) follows from the inequality 
(\ref{c39}), which is obeyed by all fractions of type II. To this end, 
let us assume that $c/d$ is the nearest fraction of $a/b$ 
in the sequence $F_N$, either from the left side, i.e.,
\bea ~~~~~~~~~~~~~~~~~~~~~~~~~~~~~~~~~~F_{N}:
~\cdots \cdots \cdots,~ \frac{c}{d} \, ,~\frac{a}{b} \, ,
\cdots \cdots \cdots ~~~~~~~~~~~~~~~~~~~~~~~(3.47a) \non \eea
or, from the right side, i.e.,
\bea ~~~~~~~~~~~~~~~~~~~~~~~~~~~~~~~~~~F_{N}:
~\cdots \cdots \cdots,~ \frac{a}{b} \, ,~\frac{c}{d} \, ,
\cdots \cdots \cdots ~~~~~~~~~~~~~~~~~~~~~~~(3.47b) \non \eea
\addtocounter{equation}{1}
Since $a/b$ is a fraction of type II, it is evident that $d \leq N-1$
(note that this statement remains valid for any $a/b$ which is not a
fraction of type I). 
By using the method of contradiction, in the following we shall show that 
$d\in \Omega_N - \Omega_{N, \phi}$ for both of the cases (3.47a) and (3.47b).
Let us first assume that $x=d$ is a zero point of the function $g(x, 
\phi)$. Thus, by using (\ref{c22}) we obtain $d=mb/a$, i.e,
\bea ~~~~~~~~m=\frac{a}{b}d\,. \non \eea
Since $\{a,b\}$ are relatively prime integers, the above equation can only be
satisfied if $d=\alpha b$, where $\alpha$ is a positive integer. However the
relation $d=\alpha b$ leads to a contradiction, since the denominators of
two consecutive fractions in a Farey sequence must be relatively prime
integers. Hence, we find that $x=d$ is not a zero point of the function 
$g(x,\phi)$. Next we assume that $x=d$ is a zero point of the function
$g(x-\bigtriangleup, \phi)$. Thus, by using (\ref{c22}) we obtain
$d-\de/a = m'b/a$, i.e., 
\bea ad - m'b = \de\,. \non \eea
Equating the (left hand side) 
l.h.s. of the above equation with that of Eq.~(\ref{c12}), we get
\bea \frac{a}{b} = \frac{t-m'}{N-d}\,. \non \eea
Since $\{a,b\}$ are relatively prime numbers, the above relation implies
that $t-m' = \beta a $ and $N-d = \beta b$, where $\beta$ is a positive
integer. However, it is easy to see that the relation $N-d = \beta b$ 
contradicts Eq.~(3.5b), which is
satisfied by the denominators of two consecutive fractions in a Farey
sequence. Hence, we find that $x=d$ is not a zero point of the function
$g(x-\bigtriangleup, \phi)$. Since $d \leq N-1$ and $x=d$ is not a zero
point of either $g(x, \phi)$ or $g(x-\bigtriangleup, \phi)$, it is
established that $d \in \Omega_N - \Omega_{N, \phi} $ for both of the cases
(3.47a) and (3.47b).

Due to Eq.~(\ref{c39}), all fractions of type II can be 
subdivided into two classes
characterized by $\de > 1$ and $\de < -1$ respectively. Let us first
consider all fractions of type II for which $\de >1$. For this case, we
assume that $c/d$ is the nearest fraction of $a/b$ (in the sequence
$F_N$) from its left side, as shown in Eq.~(3.47a). 
Consequently, by using (3.5a), we obtain 
\bea ad - bc = 1\,. \label{c48} \eea 
By choosing $m=c$ in Eq.~(\ref{c22}), one finds that $x=cb/a$ is a zero point
of the function $g(x, \phi)$. By using Eq.~(\ref{c48}), we calculate
the difference between $d$ and $cb/a$ as
\bea D\left(d,~ \frac{cb}{a} \right) = d - \frac{cb}{a} = \frac{1}{a}\,.
\label{c49} \eea
Since we are considering the case $\de > 1$, Eq.~(\ref{c21}) yields
$\bigtriangleup > 1/a$. As a result, Eq.~(\ref{c49}) leads to an 
inequality given by 
\bea D\left( d,~ \frac{cb}{a} \right) < \bigtriangleup \, , \label{c50} \eea
which shows that the zero point $x=cb/a$ of the function $g(x, \phi)$ lies
within the interval $(d-\bigtriangleup, d)$. Consequently, it is established
that $\rho(d) = 1$ for all fractions of type II with $\de > 1$. 
 
Next, let us consider all fractions of type II for which $\de < -1$. For
this case, we assume that $c/d$ is the nearest fraction of $a/b$ (in the
sequence $F_N$) from its right side, as shown in Eq.~(3.47b). 
Consequently, by using (3.5a), we obtain
\bea ad - bc = -1\,. \label{c51} \eea
By using the above relation, we obtain the distance between $cb/a$ and $d$ as 
\bea D\left( \frac{cb}{a},~ d \right) = \frac{cb}{a} - d = \frac{1}{a}\,.
\label{c52} \eea
Since we are considering the case $\de < -1$, 
Eq.~(\ref{c21}) yields $1/a < -\bigtriangleup$. 
As a result, Eq.~(\ref{c52}) leads to an inequality given by 
\bea D \left( \frac{cb}{a},~ d \right) < - \bigtriangleup \,, \label{c53} \eea
showing that the zero point $x=cb/a$ of the function $g(x, \phi)$ lies
within the interval $(d, d - \bigtriangleup )$. Consequently, it is
established that $\rho(d) = 1$ for all fractions of type II with $\de < -1$.

{}From the above analysis, it is clear that the condition (\ref{c29}) is
obeyed for all fractions of type II. Consequently, clusters of bound particles
cannot be formed for any fraction of this type.
\\
\\
{\bf Analysis of Case III}:~~ Let $\phi/\pi = a/b$ be a fraction of type III
within the set $F^{\prime}_{N-1}$. 
Since $N$ is divisible by $b$ in this case,
we may write $N = p b$, where $p ~(>1)$ is an integer. Using Eqs.~(\ref{c11}) 
and (\ref{c13}), we find that $t= a p$ and $\de = 0$ for this case. 
Substituting $\de = 0$ in Eq.~(\ref{c21}), we obtain $\bigtriangleup = 0$.
Hence, due to Eq.~(\ref{c23}), it trivially follows that $\rho(l)=0$
for all values of $l \in \Omega_N - \Omega_{N, \phi}$, i.e., 
Eq.~(\ref{c27}) is satisfied for this case. Therefore, we find that clusters 
of bound particles are formed for all fractions of type III.
According to the discussion just below Eq.~(\ref{c27}), 
the sign of $\chi$ should be chosen for this type of 
clusters of bound particles as $sgn (\chi) = {(-1)}^{ap+1}$.

{\bf Analysis of Case IV}:~~Let $\phi/\pi = a/b$ be a fraction of type IV
within the set $F^{\prime}_{N-1}$. For this case, we can write
\bea \hskip 5.24 cm b = \alpha b',~~ N=\alpha N'\,, \non \hskip 5 cm 
(3.54a,b)
\eea
\addtocounter{equation}{1}
where $\alpha, b', N'$ are some integers such that $\alpha > 1,~ b'>1$ and
$\{b', N'\}$ are relatively prime integers. Using Eqs.~(\ref{c12}) and
(3.54a,b), we obtain 
\bea \de = \alpha (aN' - tb')\,. \label{c55} \eea
Let us now assume that $aN'-tb' = 0$, which gives
\bea a = \frac{tb'}{N'} \,. \label{c56} \eea
Since $\{b', N'\}$ are relatively prime integers, the above equation would be
satisfied if $t = \beta N'$, where $\beta$ is a
positive integer. Substituting this value of $t$ in Eq.~(\ref{c56}), we obtain 
\bea a = \beta b' \,. \label{c57} \eea
Eqs.~(3.54a) and (\ref{c57}) imply that $\{a,b\}$ are not relatively
prime integers, which contradicts our basic assumption that $a/b \in
F'_{N-1}$. Thus it is established that $aN' - tb' \neq 0$. 
Since $a, N', t, b'$ are all integers, it follows that 
\bea |aN' - tb'| \geq 1 \,. \label{c58} \eea
Combining Eqs.~(\ref{c55}) and (\ref{c58}) we find that, 
similar to the case of fractions of type II, the inequality given by
\bea |\de| > 1\,, \non \eea
is satisfied for all fractions of type IV. Consequently, by carrying out the 
rest of the analysis in exactly same way as we have done in this section for 
the fractions of type II, it can be shown that all fractions of type IV
obey Eq.~(\ref{c29}). Hence, clusters of bound particles cannot be formed
for any fraction of this type.

\noi \section{Some properties of clusters of bound particles}
\renewcommand{\theequation}{4.{\arabic{equation}}}
\setcounter{equation}{0}

\medskip
 
In the previous section we have shown that, for the case of 
derivative $\de$-function Bose gas with given values of 
$\phi$ and $N$, the Bethe state (\ref{b3}) represents clusters of bound 
particles if $\phi/\pi $ is either a fraction of type I or type III
within the set $F'_{N-1}$. In this section, our aim is to find the 
number of clusters present within such a Bethe state and the sizes of these 
clusters (i.e., number of bound particles present in each of these clusters).
We would also like to investigate the behavior of these 
clusters of bound particles under small variations of the coupling constant. 

\noi \subsection{Sizes of the clusters of bound particles}

Let us first consider clusters of bound particles 
when $\phi/\pi = a/b$ is taken as any fraction of type I within the set 
$F'_{N-1}$. In section 2 we have seen that, 
to find the number and sizes of the clusters within a Bethe state,
we have to determine the set $\Omega_{N,\phi}$ for which Eq.~(3.1a)
is satisfied. Since $\{N,b\}$ are relatively prime integers 
for any fraction of type I, we can express $N$ as
\bea N=pb+r \, , \label{d1} \eea
where $1 \leq r \leq b-1 $. Hence, for the discrete variable $l$ taking values 
within the set $\Omega_N$, the zero points of the functions $\sin l\phi$ and
$\sin(N-l)\phi$ are respectively given by the sets
\bea
~~~~~~~~S_1 \equiv \{b, 2b, \cdots \cdots , pb\}, ~~~ S_2 \equiv \{ N-b, N-2b,
\cdots \cdots , N - pb \}\, . ~~~~~~~~~~(4.2a,b) \non \eea 
\addtocounter{equation}{1}
\hskip -.18 cm
Combining the sets $S_1$ and $S_2$ by using Eqs.~(\ref{d1}) and (4.2a,b), 
we obtain $\Omega_{N,\phi}$ as
\bea \Omega_{N, \phi} = S_1\cup S_2 = \{r,~b,~r+b,~2b, \cdots
\cdots,~r+(p-1)b,~pb\}\,. \label{d3} \eea
Comparing (\ref{d3}) with (\ref{b21}), and also using (\ref{b22}), 
it is easy to see that the sizes of the clusters are given by
\bea \{ \!\{r,~b-r,~r,~b-r,~\cdots \cdots,~r,~b-r,~r\}\!\}\,. \label{d4}
\eea
Hence, for any fraction of type I, the corresponding Bethe state contains 
$(p+1)$ number of clusters of size $r$ and $p$ number of clusters of size 
$(b-r)$. 
Next, by using the method of contradiction, we would like to show that these 
two possible sizes of the clusters, i.e., $r$ and $(b-r)$, must be
relatively prime integers. To this end, let us first assume that $b$ and $r$ 
are not relatively prime integers. Therefore, we can write $b$ and $r$ as 
$b=\alpha b'$ and $r=\alpha r'$, where $\alpha > 1$. Substituting these values 
of $b$ and $r$ in Eq.~(\ref{d1}), we find that \[ N = \alpha (pb' + r')\, . \]
Thus $\alpha$ is a common factor of $N$ and $b$. However, this result 
contradicts the fact that $\{N,b\}$ must be relatively prime integers for any 
fraction of type I. Hence it is established that $\{b,r\}$ are relatively 
prime integers. From this relation, it trivially follows that $\{r, b-r \}$ 
are relatively prime integers and, in particular, $r \neq b-r$.
Consequently from Eq.~(\ref{d4}) we find that, for any fraction of type I,
the corresponding Bethe state contains heterogeneous clusters of two 
different sizes. As a special case, let us consider any fraction of type I 
with denominator satisfying the relation $b > N/2$. Due to Eq.~(\ref{d1}) it 
follows that, $p=1$, $r=N-b$ and $b-r = 2b - N$ for this case. 
Hence, the corresponding Bethe state contains 
two clusters of the size $(N-b)$ and one cluster of the size $(2b - N)$.

Next, we consider the clusters of bound particles corresponding to any 
fraction of type III. In this case $N$ can be written as $N=pb$, 
where $p$ is an integer greater than one. Consequently, for the variable $l$ 
taking value within the set $\Omega_N$, the zero points of the functions 
$\sin l\phi$ and $\sin(N-l)\phi$ coincide with each other and yield 
$\Omega_{N, \phi}$ as 
\bea \Omega_{N, \phi}= \{ b,~2b,~3b,\cdots \cdots ,(p-1)b \}\,. \label{d5} \eea
Hence $p$ number of clusters are formed in this case.
Comparing (\ref{d5}) with (\ref{b21}), and also using (\ref{b22}), 
it is easy to see that each cluster of this type has the size $b$. 
In other words, the corresponding Bethe state (\ref{b3}) contains $N/b$ number
of homogeneous clusters, each of which is made of $b$ number of bound 
particles.

In Table 1 we show all the fractional values of $\phi /\pi$ for which clusters
of bound particles exist within the range of $N$ given by $4 \leq N \leq 10$,
the types of these fractions, the values of $\de$ for these fractions, and
the sizes of the corresponding clusters using the notation of Eq.~(\ref{b22}).

\vspace{0.4cm}
\begin{table}[htb]
\begin{center}
\begin{tabular}{|c|c|c|r|c|} \hline
$N$ & Value of $\phi/\pi$ & Type & Value of $\de$ & Size of the clusters \\ 
\hline
4 &$1/3$ & I & 1~~~~~~ & $\{ \! \{ 1,2,1 \} \! \}$\\
5 & $1/4$ & I & 1~~~~~~ & $\{ \!\{1,3,1\}\!\}$\\ 
5 & $1/3$ & I & $-1~~~~~~$ & $\{\!\{2,1,2\}\!\}$\\
6 & $1/5$ & I & 1~~~~~~ & $\{\!\{1,4,1 \}\!\}$\\ 
6 & $1/3$ & III & 0~~~~~~ & $\{\!\{3,3\}\!\}$ \\
7 & $1/6$ & I & 1~~~~~~ & $\{\!\{1, 5, 1 \}\!\}$\\ 
7 & $1/4$ & I & $-1~~~~~~$ & $\{\!\{3,1,3 \}\!\}$\\
7 & $1/3$ & I & 1~~~~~~ & $\{\!\{1,2,1,2,1 \}\!\}$ \\
7 & $2/5$ & I & $-1~~~~~~$ & $\{\!\{2,3,2 \}\!\}$\\
8 & $1/7$ & I & 1~~~~~~ & $\{\!\{1,6,1\}\!\}$\\ 
8 & $1/4$ & III & 0~~~~~~ & $\{\!\{ 4, 4 \}\!\}$ \\
8 & $1/3$ & I & $-1~~~~~~$ & $\{\!\{ 2, 1, 2, 1, 2 \}\!\}$\\
8 & $2/5$ & I & 1~~~~~~ & $\{\!\{3, 2, 3 \}\!\}$ \\
9 & $1/8$ & I & 1~~~~~~ & $\{\!\{1, 7, 1 \}\!\}$\\ 
9 & $1/5$ & I & $-1~~~~~~$ & $\{\!\{ 4, 1, 4 \}\!\}$\\
9 & $1/4$ & I & 1~~~~~~ & $\{\!\{ 1, 3, 1, 3, 1 \}\!\}$ \\
9 & $1/3$ & III & 0~~~~~~ & $\{\!\{3, 3, 3\}\!\}$ \\
9 & $3/7$ & I & $-1~~~~~~$ & $\{\!\{ 2, 5, 2 \}\!\}$\\ 
10 & $1/9$ & I & 1~~~~~~ & $\{\!\{ 1,8,1 \}\!\}$ \\
10 & $1/5$ & III & 0~~~~~~ & $\{ \! \{ 5, 5 \}\!\}$ \\
10 & $2/7$ & I & $-1~~~~~~$ & $\{\!\{ 3,4,3 \}\!\}$ \\ 
10 & $1/3$ & I & 1~~~~~~ & $\{\!\{ 1, 2, 1, 2, 1, 2, 1 \}\!\}$ \\
10 & $2/5$ & III & 0~~~~~~ & $\{\!\{ 5, 5 \}\!\}$ \\ \hline
\end{tabular}
\end{center} 
\caption{The fractional values of $\phi /\pi$ for which clusters of bound 
particles exist for $4 \leq N \leq 10$, the types of these fractions, the
values of $\de$ for fractions of type I, and the sizes of the corresponding 
clusters are shown.} 
\end{table}

\vspace{0.2cm}

\noi \subsection{Stability of clusters of bound particles and binding energies}

We want to explore here how the nature of clusters of bound particles
change under slight variations of the coupling constant.
For any given value of $N$ ($\geq 4$), we first choose a value of the 
coupling constant $\phi$ within the range $0<\phi<\frac{\pi}{2}$
such that clusters of bound particles are formed.
If one increases or decreases this value of $\phi$ by an infinitesimal amount, 
it is obvious that all inequalities in Eq.~(3.1b) continue to be satisfied, 
but all equalities in Eq.~(3.1a) are transformed into some inequalities.
Therefore, clusters of bound particles cease to exist even for a 
very small change of the coupling constant. One of the following 
two different cases can occur in such a situation. In the first case, 
at least one of the equalities in Eq.~(3.1a) is transformed 
into an inequality of the form 
\beq \chi ~\sin (l \phi)~\sin [(N-l) \phi] ~<~ 0 \, . \label{d6} \eeq
It is evident that, the probability density of the corresponding
Bethe state (\ref{b3}) would diverge if at least 
one of the particle coordinates
tends towards infinity (keeping the centre of mass coordinate fixed).
As a result, this Bethe state becomes ill-defined and disappears from the 
Hilbert space of the Hamiltonian (\ref{a2}) of derivative $\de$-function 
Bose gas. So we may say that clusters of bound particles become unstable
in this case. Let us now consider the second case, for which all of equalities
in Eq.~(3.1a) are transformed into inequalities of the form 
\beq \chi ~\sin (l \phi)~\sin [(N-l) \phi] ~>~ 0\, , \label{d7} \eeq
due to an infinitesimal change of the coupling constant $\phi$. It is evident 
that, for this case, Eqs.~(3.1a,b) are transformed to Eq.~(\ref{b16}) within 
the range of $\phi$ given by $0<\phi<\frac{\pi}{2}$.
As a result, clusters of bound particles merge with each other and 
produce a localized bound state containing only one cluster of particles.
Therefore, in this case, we may say that clusters of bound particles turn 
into a localized bound state with only one cluster. 

In this context, we note that the first and second derivatives
of the function $f(l, N, \phi)$ in Eq.~(\ref{c17}) are given by
\bea
&&~~~~~~~~~~\frac{\partial f}{\partial \phi} = \frac{\chi N}{2} \sin (N\phi) -
\frac{\chi (N-2l)}{2} \sin (N-2l)\phi \,, 
~~~~~~~~~~~~~~~~~~~~~~~~~~~~~~~(4.8a)\non \\ 
&&~~~~~~~~~~\frac{\partial^2 f}{\partial \phi^2} = \frac{\chi N^2}{2}
 \cos (N\phi) - \frac{\chi {(N - 2l)}^2}{2} \cos (N-2l) \phi\,. \non
~~~~~~~~~~~~~~~~~~~~~~~~~~~~(4.8b) \eea
\addtocounter{equation}{1}
By calculating these derivatives at some value of 
$\phi$ for which clusters of bound states exist and at a value of $l$
within the set $\Omega_{N, \phi}$, one can find whether 
the corresponding equality in Eq.~(3.1a) is transformed into an inequality 
of type (\ref{d6}) or (\ref{d7}) 
due to an infinitesimal change in the coupling constant. In the following,
we shall use this procedure to determine the nature of 
clusters of bound particles for any given values of $\phi$ and $N$. 

Let us first consider clusters of bound particles
when $\phi/\pi = a/b$ is taken as any fraction of type III.
As mentioned earlier, in this case $N$ can be written as $N=pb$ 
(where $p$ is an integer) and from Eq.~(\ref{c13}) it follows that $t=ap$.
Moreover, due to Eq.~(\ref{d5}), $l$ should be taken to be of the form $l=mb$,
where $m$ is an integer. For the values of $N$, $\phi$ and $l$ given by
$N=pb$, $\phi=\pi a/b$ and $l=mb$, we get $\sin (N\phi) = \sin (\pi a p) = 0\,$ 
and $~\sin (N - 2l)\phi = \sin (N\phi - 2\pi ma) = 0$.
Thus, by using Eq.~(4.8a), we obtain
\bea \frac{\partial f}{\partial \phi} \,
{\put(0,-12.5){\line(0,1){29}}}_{\, \phi=\frac{\pi a}{b},\, l=mb} \, = \, 0\, ,
\label{d9} \eea 
which shows that $\phi=\pi a/b$ is an extremum point of the function 
$f(l, N, \phi)$. To determine the nature of this extremum point, it is 
necessary to calculate the second derivative of $f(l, N, \phi)$. To this end 
we recall that, according to our discussion just after Eq.~(\ref{c27}),
the value of $\chi$ should be chosen such that the inequality ${(-1)}^t\chi < 
0$ is satisfied. Since $t=ap$ for all fractions of type III, we can rewrite 
the above mentioned inequality as
\beq {(-1)}^{ap}\chi < 0. \label{d10} \eeq
Moreover, for the case $N=pb$, $\phi=\pi a/b$ and $l=mb$, we obtain 
$\cos (N\phi) =\cos(N-2l)\phi ={(-1)}^{ap}$. Substituting these values of 
$\cos (N\phi)$ and $ \cos(N-2l)\phi$ in Eq.~(4.8b), and subsequently using 
the inequality (\ref{d10}), we find the relation
\bea \frac{\partial^2 f}{\partial \phi^2} \, 
{\put(0,-12.5){\line(0,1){29}}}_{\, \phi=\frac{\pi a}{b},\, l=mb}
\, = \, 2l(N-l)\cdot \chi {(-1)}^{ap} < 0 \, , \label{d11} \eea
which shows that the function $f(l, N,\phi)$ has a maxima at $\phi=\pi a/b$. 
Consequently, if one moves slightly away from the point $\phi=\pi a/b$ (in 
positive or negative direction), the corresponding equality in Eq.~(3.1a) is 
transformed into an inequality of the form (\ref{d6}). 
Therefore, we find that the clusters of bound particles are unstable around
the point $\phi=\pi a/b$, where $a/b$ is a fraction of type III. 

Next, we consider clusters of bound particles corresponding to any fraction of 
type I. Due to Eq.~(\ref{c34}), it is possible to divide
all fractions of type I into two subclasses,
corresponding to the value of $\de$ given by $+1$ and $-1$ respectively.
We shall see below that clusters of bound particles related to these 
two subclasses behave differently under the variation of the coupling constant.
For any fraction of type I, 
the union of the sets $S_1$ and $S_2$ given 
in Eqs.~(4.2a,b) yields the set $\Omega_{N, \phi}$.
Therefore, the values of the corresponding $l$ should be taken either as $l=mb$
or as $l=N-mb$, where $m$ is an integer. Let us first consider the case when 
$l=mb$. Since $\sin (N-2l)\phi = \sin (N\phi)$ in this case, by using 
Eq.~(\ref{c12}) we obtain 
\beq \sin (N-2l)\phi = \sin (N\phi) 
= {(-1)}^t \sin \left(\frac{\pi \de}{b}\right)\,. \label{d12} \eeq
Similarly, for the case $l=N-mb$, we obtain
\beq - \sin (N-2l)\phi = \sin (N\phi) 
= {(-1)}^t \sin \left(\frac{\pi \de}{b}\right)\,. \label{d13} \eeq
Substituting the values of $\sin (N\phi)$ and $\sin (N-2l)\phi$ given in 
Eqs.~(\ref{d12}) and (\ref{d13}) to Eq.~(4.8a) for the cases $l=mb$ and 
$l=N-mb$ respectively, and also using the inequality ${(-1)}^t\chi <0$, we get
\bea \frac{\partial f}{\partial \phi} \,
{\put(0,-12.5){\line(0,1){29}}}_{\, \phi=\frac{\pi a}{b},\, l=mb} =
\frac{\partial f}{\partial \phi} \,
{\put(0,-12.5){\line(0,1){29}}}_{\, \phi=\frac{\pi a}{b},\, l=N-mb} =
mb \cdot \chi {(-1)}^t \cdot \sin \left(\frac{\pi\de}{b} 
\right) = \left \{ \begin{array}{l}
< 0 \rm{~for~} \de = 1 \, , \\
> 0 \rm{~for~} \de = -1\,. \end{array} \right. \label{d14} \eea
Hence we find that, for the case of fractions of type I with $\de = 1$, 
clusters of bound particles would become unstable due to a slight increase of 
the value of $\phi$. On the other hand, such clusters of bound particles 
would transmute to a localized bound state containing only one cluster if 
the value of $\phi$ is slightly decreased. For the fractions of type I with 
$\de = -1$, clusters of bound particles would behave in exactly reversed 
order if one slightly increases or decreases the value of $\phi$.

It is interesting to calculate the binding energies of the clustered
states that we have been considering so far. The binding energy $E_B$ of a 
clustered state is defined as follows. The momentum $P$ and energy $E$ of 
such a state were presented in Eqs.~(\ref{b23}) and (\ref{b24}).
Let us now consider a state of $N$ free particles which are all moving with
the same momentum $p$ so that the total momentum $Np$ is equal to the 
expression in Eq.~(\ref{b23}). Clearly,
\beq p ~=~ \hbar \chi ~\frac{\sin (N\phi)}{N \sin \phi} ~. \eeq
The energy of this free particle state is given by
\beq E_0 ~=~ N p^2 ~=~ \frac{\hbar^2 \chi^2 \sin^2 (N \phi)}{N \sin^2 
(\phi)} ~. \eeq
The binding energy $E_B$ corresponding to clusters of bound particles 
is defined as the difference of the energies of the
free particle state and the clustered state, i.e.,
\bea ~~~~~~~~E_B &\equiv& E_0 ~-~ E ~=~ \frac{\hbar^2 \chi \sin (N \phi)}{N 
\sin^2 (\phi) \cos \phi} ~h(N,\phi) ~, ~~~~~~~~~~~~~~~~~~~~~~~~~~~~(4.17a)
\non \\
~~~~~~~~h(N,\phi) &=& \chi ~[\sin (N \phi) \cos \phi ~-~ N \cos (N \phi) 
\sin \phi]. ~~~~~~~~~~~~~~~~~~~~~~~~~~~~~~(4.17b) \non \label{hnp} \eea
\addtocounter{equation}{1}
It should be noted that Eqs.~(4.17a,b) also give the binding energies of 
localized bound states for the case of the derivative $\de$-function Bose gas 
\cite{BBS03,BBS04a}. We now consider the two kinds of clusters of bound 
particles in which $\phi/\pi =a/b$ corresponds to fractions of types I and III
within the sequence $F'_{N-1}$.
For the case of fractions of type III, $N$ is a multiple of $b$; hence
$\sin (N \phi) = 0$. Thus the momentum $p$ and the energies $E$ and $E_0$ are 
all equal to zero in this case, and $E_B = 0$ also. We now turn to fractions
of type I. In Refs. \cite{BBS03,BBS04a}, it was shown that $h (N,\phi) > 0$ for
all these values of $\phi$. Eq.~(4.17a) then implies that $E_B$ has the same 
sign as $\chi \sin (N\phi)$. Combining Eqs.~(\ref{d12}-\ref{d14}), we see that
$\chi \sin (N\phi)$ and hence $E_B$ are negative for $\de =1$ (i.e., if
$(aN-1)/b$ is an integer) and are positive for $\de = -1$ (i.e., if
$(aN+1)/b$ is an integer).

In our earlier works it was shown that, for any given value of $N \geq 4$,
the derivative $\de$-function Bose gas allows bound states in only certain 
non-overlapping ranges of the coupling constant $\phi$ called `bands'
(the union of which yields a proper subset of the full range of $\phi$),
and the location of these bands
can be determined exactly \cite{BBS03,BBS04a}.
Applying the results obtained in this section, we shall now present
an alternative method of computing the location of these bands
within the full interval $(0,\pi/2)$ for positive values of $\phi$. 
Since fractions of type I only lead to clusters of bound particles 
which transmute to bound states with a single cluster through 
infinitesimal changes of $\phi$ in appropriate directions, it is evident 
that any fraction of type I or $0$ or $1/2$ must appear at the two 
edges of any continuous range of $\phi/\pi$ within which localized bound 
states are formed. Therefore, we include the numbers $0$ and $1/2$ 
along with the fractions of type I (corresponding to the set $F'_{N-1}$) 
and call this new set as type ${\rm I}'$. Next, we denote two fractions of 
type ${\rm I}'$ as $a_i/b_i$ and $a_i'/b_i'$, and define them as `conjugate 
pair' if they appear within the Farey sequence $F_{N}$ as 
\beq F_N : ~\cdots \cdots \frac{a_i}{b_i} ~\frac{n_i}{N} ~ \frac{a_i'}{b_i'}
\cdots \cdots \, , \label{d15} \eeq
where $n_i/N \in F''_N $. Let us also assume that all possible
conjugate pairs of this type can be obtained by varying the index $i$
within the range $i \in \{ 1,2, \cdots , \sigma(N) \}$. 
Clearly, for any $N \geq 4$, conjugate pairs defined 
in Eq.~(\ref{d15}) can be related to three different cases: 
(i) $a_i/b_i$ and $a_i'/b_i'$ are both fractions of type I, (ii)
$ a_i/b_i=0$ and $a_i'/b_i'$ is a fraction of type I, 
and (iii) $a_i'/b_i'=1/2$ and $a_i/b_i$ is a fraction of type I.
Let us first consider any conjugate pair corresponding to the case (i). 
Using Eqs.~(3.5a) and (\ref{c12}) along with the 
relation $t=n$, it is easy to see that $\de= -1$ for $a_i/b_i$
and $\de= 1$ for $a_i'/b_i'$. Therefore, from the discussion of the 
previous paragraph it follows that clusters of bound particles 
would be transformed to a localized bound state only if one
increases the value of $\phi/\pi$ from $a_i/b_i$ or 
decreases the value of $\phi/\pi$ from $a_i'/b_i'$
by an infinitesimal amount. Moreover it should be noted that, since 
$a_i/b_i$ and $a_i'/b_i'$ are consecutive fractions within the 
set $F'_{N-1}$, the sign of the function $f(l, N, \phi)$
cannot flip even for any finite amount of change of $\phi/\pi$ 
within the interval $(a_i/b_i, a_i'/b_i')$. Consequently, for any conjugate
pair corresponding to case (i), localized bound states
continue to exist if $\phi/\pi$ takes any value within the range 
$(a_i/b_i, a_i'/b_i')$. By repeating similar arguments, one can reach the 
same conclusion for any conjugate pair associated with cases (ii) or (iii). 
Thus we find that, localized bound states can be constructed for any value 
of $\phi$ within the non-overlapping ranges or bands given by
\beq {\phi} \in \left( \frac{\pi a_i}{b_i} \, ,\, \frac{\pi a_i'}{b_i'}
\right) , \label{d16} \eeq
where $i \in \{ 1,2, \cdots , \sigma(N) \}$. Due to Eq.~(\ref{d15})
it is evident that there exists an one-to-one correspondence between the 
conjugate pairs formed by fractions of type ${\rm I}'$ and the elements of 
the set $F''_N$. Therefore, the elements of $F''_N$ can be used to uniquely 
characterize the bands appearing in Eq.~(\ref{d16}). Moreover, it is easy to 
check that Eq.~(\ref{d15})
provides the only possible way of relating two fractions like $a/b$ and $c/d$
of type ${\rm I}'$, such that localized bound states are formed for any value 
of $\phi/\pi$ within the range $(a/b,c/d)$. Hence, localized bound states can 
only be constructed within the ranges of $\phi$ given in Eq.~(\ref{d16}).

For an illustration of the idea of bands, and the stability of clusters
of bound particles under a variation of $\phi$ and their binding energy 
discussed above, we calculate the binding energies of localized bound states
within all bands for $N=7$, 8 and 9 by using Eqs.~(4.17a,b), and we show these
binding energies as a function of $\phi/\pi$ in Fig.~1. 
Comparing this figure with the appropriate entries in Table 1, we see that
the values of $\phi/\pi$ which are fractions of type III do not lie in 
any bands, while fractions of type I lie at the end points of the bands. 
Furthermore, fractions of type I with $\de = 1 ~(-1)$ lie at the right (left) 
end of the bands and have $E_B < 0 ~(> 0)$ respectively.

\begin{figure}[htb]
\centering 
\hskip .25 cm 
\includegraphics[scale=1]{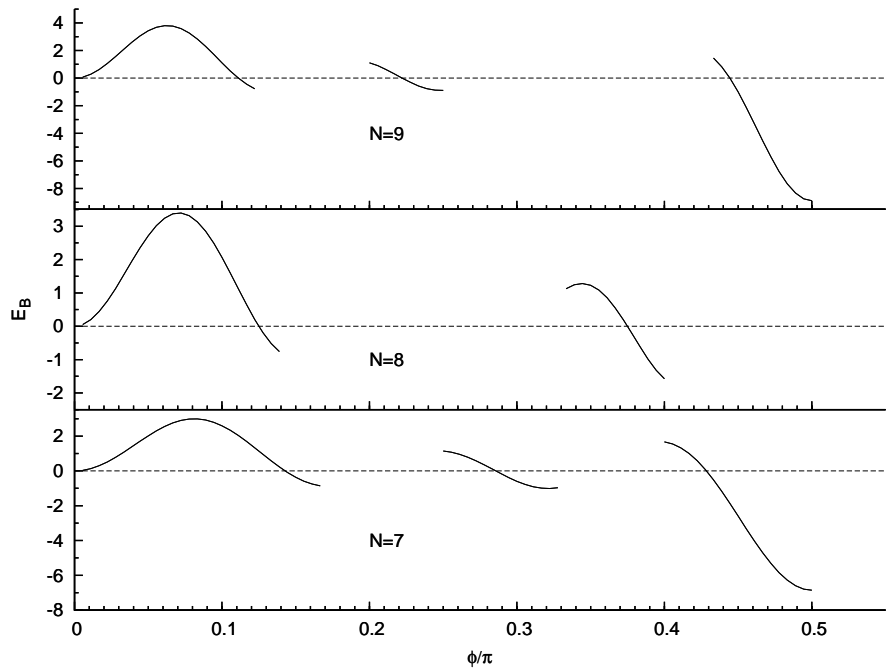}
\caption{The binding energy $E_B$ of the localized bound states as a 
function of $\phi/\pi$ for $N=7$, 8 and 9.}
\medskip
\end{figure}

\noi \section{Conclusion}
\renewcommand{\theequation}{5.{\arabic{equation}}}
\setcounter{equation}{0}

\medskip
 
In this article, we have explored how 
clusters of bound particles can be constructed in the simplest possible
way for the case of an exactly solvable derivative $\de$-function Bose gas. 
To this end, we consider a sufficient condition for which Bethe states of 
the form in (\ref{b3}) would lead to clusters of bound particles.
This sufficient condition for obtaining clusters of bound particles has not 
attracted much attention earlier, probably because it does
not yield any solution at all for the case of the $\de$-function Bose gas.
However we find that, for the case of the derivative $\de$-function Bose gas,
this sufficient condition can be satisfied by taking the quasi-momenta of the 
corresponding Bethe state to be equidistant points on a {\it single}
circle having its centre at the origin of the complex momentum plane. 
Furthermore, the coupling constant ($\phi$) and the total number of particles 
($N$) of this derivative $\de$-function Bose gas must satisfy the relations in 
(3.1a,b). For any given $N\geq 4$, it is found that Eq.~(3.1a) is satisfied
if $\phi/\pi$ takes any value within the set $F'_{N-1}$, which is 
a subset of the Farey sequence $F_{N-1}$. Then we classify all fractions 
belonging to the set $F'_{N-1}$ into four types. It turns out that fractions
of types I and III belonging to $F'_{N-1}$ satisfy the remaining relation 
(3.1b), while type II and type IV fractions do not satisfy (3.1b).
Consequently, clusters of bound particles can be constructed for the 
derivative $\de$-function Bose gas only for special values of $\phi/\pi$ 
given by the fractions of types I and III within the set $F'_{N-1}$. 

We also computed the sizes of the above mentioned clusters of bound particles,
i.e., the number of particles present within each of these clusters. We find 
that any fraction of type I within the set $F'_{N-1}$ leads to heterogeneous 
clusters of bound particles having two different sizes. On the other hand, 
any fraction of type III within the set $F'_{N-1}$
leads to homogeneous clusters of bound particles having only one size.
Interestingly, clusters of bound particles associated with fractions of 
type I and type III transform in rather different ways under a small
variation of the coupling constant. For example, it is found that
clusters of bound particles associated with fractions of type III 
cease to exist for any small change of the coupling constant. On the other 
hand, clusters of bound particles corresponding to fractions of type I turn
into localized bound states consisting of a single cluster if the value of 
the coupling constant is slightly increased or decreased. Finally, we find 
that clusters of bound particles associated with fractions of type III have 
zero binding energy, while clusters associated with fractions of type I have 
either positive or negative binding energy. 

Even though in this paper we have analyzed a 
particular type of sufficient condition for constructing clusters
of bound particles in the case of the derivative $\de$-function Bose gas, in 
future it would be interesting to explore other possible ways of constructing 
clusters of bound particles for this exactly solvable system. We have 
previously mentioned that, clusters of bound particles are constructed for 
the case of the $\de$-function Bose gas by taking the corresponding 
quasi-momenta to be discrete points lying on several parallel straight lines 
in the complex momentum plane. In analogy with this case, one may try to 
construct clusters of bound particles for the present case by assuming that 
the corresponding quasi-momenta lie on several concentric circles in the 
complex momentum plane. However it should be noted that, for the case of the 
derivative $\de$-function Bose gas with any given value of the coupling 
constant, localized bound states can appear only for some particular 
values of the particle number \cite{BBS03,BBS04a}. Hence, for constructing 
clusters of bound particles in this case, the number of quasi-momenta on each 
concentric circle must be taken from the subset of such allowed 
particle numbers. Due to this constraint on the numbers of quasi-momenta,
it remains a rather challenging problem to construct all possible clusters 
of bound particles in the case of the derivative $\de$-function Bose gas.

\vspace{1cm}
\noi {\bf Acknowledgments}
\vspace{.3cm} 

\noi 
B.B.M. thanks the Abdus Salam International Centre for Theoretical Physics for 
a Senior Associateship, which partially supported this work. D.S. thanks the 
Department of Science and Technology, India for financial support through 
the grant SR/S2/JCB-44/2010.

\newpage
\leftline {\large \bf Appendix A}
\baselineskip=14pt
\medskip
\noindent
In this Appendix, we shall explicitly derive the relation (\ref{c15}). 
At first, we shall show that if $a/b\in F'_{N-1}$ satisfies the condition
\bea ~~~~~~~~~~~~~~~~~~~~~~~~~~~~~~~~~~~~~~~~~~\frac{a}{b}~ \leq ~
\frac{1}{2} - \frac{1}{2N} \, ,~~~~~~~~~~~~~~~~~~~~~~~~~~~~~~~~~~~~~~~~~~~
(A1) \non \eea
then $t/N$ will satisfy the relation (\ref{c15}). To this end, we rewrite
Eq.~(\ref{c12}) as
\bea ~~~~~~~~~~~~~~~~~~~~~~~~~~~~~~~~~~~~~~~~~\frac{t}{N} = \frac{a}{b} -
\frac{\de}{bN}\,.
~~~~~~~~~~~~~~~~~~~~~~~~~~~~~~~~~~~~~~~~~~~~~~(A2) \non \eea
Eqs.~(A1) and (A2) imply that 
\bea ~~~~~~~~~~~~~~~~~~~~~~~~~~~~~~~~~~~~~~~
\frac{t}{N} ~\leq ~\frac{1}{2} - \frac{1}{2N} -
\frac{\de}{bN}\,.~~~~~~~~~~~~~~~~~~~~~~~~~~~~~~~~~~~~~~(A3) \non \eea
For the case $\de \geq 0$, the relation (\ref{c15}) directly follows from
Eq.~(A3). For the case $\de < 0$, Eq.~(\ref{c13}) yields
\bea ~~~~~~~~~~~~~~~~~~~~~~~~~~~~~~~~~~~~~~~~~~~-\frac{\de}{b} <
\frac{1}{2}\,.~~~~~~~~~~~~~~~~~~~~~~~~~~~~~~~~~~~~~~~~~~~~~ ~~~~~(A4)
\non \eea
Combining (A3) and (A4), it can be easily seen that the relation (\ref{c15})
is satisfied for $\de < 0$. Thus we are able to prove that the relation
(\ref{c15}) follows from the condition (A1) for all possible values of $\de$.

Next, for the purpose of proving the validity of condition (A1), let us
consider the cases of even and odd $N$ separately. For the case of even values
of $N$, we may write $N=2m$. Using the properties (3.5a,b) 
of Farey sequence, we find that 
\bea F_{2m-1}:~ \cdots \cdots \cdots ~\frac{m-1}{2m-1}~~\frac{1}{2} ~\cdots
\cdots \cdots\,, \non \eea
i.e., the fraction nearest to $1/2$ from the left side of the
sequence $F_{2m-1}$ is given by $(m-1)/(2m-1)$. Consequently, any fraction 
$a/b \in F'_{2m-1}$ satisfies the relation 
\bea ~~~~~~~~~~~~~~~~~~~~~~~~~~~~~~~~~~~~~~~~~~\frac{a}{b}~ \leq ~
\frac{m-1}{2m-1}\,.
~~~~~~~~~~~~~~~~~~~~~~~~~~~~~~~~~~~~~~~~~~~~~(A5) \non \eea 
Using (A5), we easily find that the condition (A1) with 
$N=2m$ is satisfied by any $a/b \in F'_{2m-1}$.
Next, we consider the case of odd values of $N$ and write $N=2m+1$.
Using Eqs.(3.5a,b), it is easy to check that there are two consecutive 
fractions in the Farey sequence $F_{2m}$ given by 
\bea F_{2m}: ~\cdots \cdots \cdots~ \frac{m-1}{2m-1}~~\frac{1}{2}~ \cdots
\cdots \cdots \, . \non \eea
Consequently, any fraction $a/b \in F'_{2m}$ also satisfies the relation (A5). 
Using (A5), we find that the condition (A1) with $N=2m+1$ is satisfied by any 
$a/b \in F'_{2m}$. Thus it is established that the condition (A1) is satisfied 
by any $a/b \in F'_{N-1}$ for both even and odd values of $N$. 

\newpage
\leftline {\large \bf Appendix B}
\baselineskip=14pt
\medskip
\noindent
Here our aim is to show that for any $a/b \in F'_{N-1}$, there exists at
least one $l \in \Omega_N - \Omega_{N, \phi}$ such that $\rho(l) = 0$. To
this end, we shall consider the cases $\de \geq 0$ and $\de < 0$ separately. 
\vskip .2 cm
\noindent{\bf Case (i)~}
Let us first consider the case $\de \geq 0$, for which $\bigtriangleup
\geq 0$. Using Eqs. (3.3a) and (\ref{c14}), we obtain the relation 
\bea ~~~~~~~~~~~~~~~~~~~~~~~~~~~~~~~~~~~~~~~~~~~~~~~~0 < a+\de < b\,.
~~~~~~~~~~~~~~~~~~~~~~~~~~~~~~~~~~~~~~~(B1) \non \eea
By using the method of contradiction, at first we shall show that $(b-1)\in
\Omega_N - \Omega_{N, \phi}$. Let us first assume that $x = b-1$ is a zero 
point of the function $g(x, \phi)$. Then by using Eq.~(\ref{c22}) we obtain
$b-1=mb/a$, which yields
\bea ~~~~~~~~~~\frac{a}{b} = a - m\,. \non \eea
However, the above equation leads to a contradiction because its l.h.s. is a 
proper fraction and its right hand side (r.h.s.) is an integer. Thus $x = b-1$
cannot be a zero point of the function $g(x, \phi)$. Next, we assume that
 $x = b-1$ is a zero point of the function $g(x - \bigtriangleup, \phi)$. Then
by using Eq.~(\ref{c22}) we obtain $b-1-\de/a = m'b/a$ , which yields
\bea ~~~~~~~~~~~~~~\frac{a+\de}{b} = a - m'\,. \non \eea
Again the above equation leads to a contradiction because its r.h.s.
is an integer and its l.h.s. is a proper fraction due to
Eq.~(B1). Thus $x = b-1$ cannot be a zero point of the function $g(x -
\bigtriangleup, \phi)$. Since $(b-1) < (N-1)$ and $x = b-1$ is not a zero 
point of either $g(x, \phi)$ or $g(x - \bigtriangleup, \phi)$, 
it follows that $(b-1) \in \Omega_N - \Omega_{N, \phi}$.

Next, our aim is to show that $\rho(l)=0$ for the choice $l=b-1$. To this end, 
let us assume that $x = m''b/a$ is a zero point of the function $g(x,
 \phi)$ and $m''b/a < (b-1)$. Since the difference between $(b-1)$ and
$m''b/a$ may be written as 
\bea
~~~~~~~~~~~~~~~~~~~~~~~~~~~~~~~~~~~D\left(b-1, \frac{m''b}{a}\right) = -1 + 
\frac{b}{a} (a- m'')\, ,~~~~~~~~~~~~~~~~~~~~~~~~~(B2) \non \eea
it is clear that $m''<a$. 
Using Eqs.~(3.3a) and (B2), it is easy to see that $D(b-1, m''b/a)$ yields
the minimum positive value for the choice $m'' = a-1$. 
Furthermore, by using Eqs.~(B1) and (\ref{c21}), we find that this minimum
positive value of $D(b-1, m''b/a)$ satisfies the relation
\bea ~~~~~~~~~~~~~~~~~~~~~~~~~~~~~~~~~~~~\left.D\left(b-1,\frac{m''b}{a}\right)
\right|_{m''=a-1} = \frac{b-a}{a} > \bigtriangleup\,.
~~~~~~~~~~~~~~~~~~~~~ \non \eea
{}From the above equation it follows that, there exists no zero point of the
function $g(x, \phi)$ in the variable $x$ within the interval $(b-1-
\bigtriangleup, b-1)$. Thus, for any $a/b \in F'_{N-1}$ with $\de \geq 0$, we 
are able to establish that $\rho(b-1)=0$.

\vskip .2 cm
\noindent{\bf Case (ii)~}
Let us consider the case $\de < 0$, for which $\bigtriangleup < 0$.
Using Eqs.~(3.3a) and (\ref{c14}), we obtain the relation 
\bea
~~~~~~~~~~~~~~~~~~~~~~~~~~~~~~~~~~~~~~~~~~0 < a-\de <
b\,.~~~~~~~~~~~~~~~~~~~~~~~~~~~~~~~~~~~~~~~~~~~~~(B3) \non \eea
By using the method of contradiction, in the following we shall show that
$(b+1) \in \Omega_N-\Omega_{N, \phi}$, i.e, $(b+1) \leq (N-1)$ and $x = (b+1)$
is not a zero point of either $g(x, \phi)$ or $g(x -
\bigtriangleup, \phi)$. For this purpose it may be noted that , since
$a/b \in F'_{N-1}$, one can in principle choose $b=N-1$, which contradicts the 
condition $(b+1) \leq (N-1)$. However, by applying Eqs.~(\ref{c11}) and 
(\ref{c13}) for this case, we obtain $[Na/b] = a $ and $\de = a > 0$. Thus 
the choice $b=N-1$ is not compatible with the present case associated with 
negative values of $\de$.
Next we assume that $x = b+1$ is a zero point of the function $g(x, \phi)$. 
Then by using Eq.~(\ref{c22}) we obtain $b+1 = mb/a$, which yields
\bea ~~~~~~~\frac{a}{b} = m-a\,. \non \eea
However, the above equation leads to a contradiction because its l.h.s.
is a proper fraction and its r.h.s. is an integer. Thus $x
= b+1$ cannot be a zero point of the function $g(x, \phi)$. Next, we
assume that $x = b+1$ is a zero point of the function $g(x -
\bigtriangleup, \phi)$. Then by using Eq.~(\ref{c22}) we obtain
$b+1-\de/a = m'b/a$, which yields
\bea \frac{a - \de}{b} = m' - a\,. \non \eea
Again, the above equation leads to a contradiction because its r.h.s.
is an integer and its l.h.s. is a proper fraction due to
Eq.~(B3). Thus $x = b+1$ cannot be a zero point of the function
$g(x-\bigtriangleup, \phi)$. As a result, we find that
 $(b+1) \in \Omega_N - \Omega_{N, \phi}$
for any fraction $a/b \in F'_{N-1}$ which leads to $\de < 0$. 

Next, our aim is to show that $\rho(l) = 0$ for the choice $l= b+1$. To this 
end, let us assume that $x = m''b/a$ is a zero point of the function $g(x,
\phi)$ and $m''b/a >(b+1)$. Since the difference between $m''b/a$ 
and $(b+1)$ may be written as 
\bea ~~~~~~~~~~~~~~~~~~~~~~~~~~~~~~~~~~D\left(\frac{m''b}{a} \, , b+1\right) =
-1 + \frac{b}{a}(m''-a)\, ,~~~~~~~~~~~~~~~~~~~~~~~~~(B4) \non \eea
it is evident that $m''>a$. With the help of 
Eqs.~(3.3a) and (B4), it is easy to check that $D(m''b/a, b+1)$ yields the
minimum positive value for the choice $m'' = a+1$. Furthermore,
by using Eqs.~(B3) and (\ref{c21}), 
we find that this minimum positive value of $D(m''b/a \, , b+1)$
satisfies the relation 
\bea ~~~~~~~~~~~~~~~~~~~~~~~~~~~~~~~~~~\left.D\left(\frac{m''b}{a},
b+1\right)\right|_{m''= a+1} = \frac{b-a}{a} > -\bigtriangleup
\,.~~~~~~~~~~~~~~~~~~~(B5) \non \eea
Hence, it follows that there exists no zero point of the function $g(x, \phi)$
in the variable $x$ within the interval $(b+1, b+1-\bigtriangleup)$.
Thus for any $a/b \in F'_{N-1}$ with $\de < 0$, we are able to establish
that $\rho(b+1) = 0$.

\end{document}